\documentstyle[prb,epsf,multicol,aps]{revtex}

\begin{document}

%\draft

\tighten

%\preprint{\today}
%\preprint{MA/UC3M/02/95}

\title{{\rm Phys.\ Rev.\ B, submitted}
\hfill {MA/UC3M/{\rm 02}/{\rm 1995}}\\[2mm]
Transport properties of nonlinear double-barrier structures}

\author{Enrique Diez and Angel S\'{a}nchez}

\address{Escuela Polit\'{e}cnica Superior, Universidad
Carlos III de Madrid, C./ Butarque 15, % \\
E-28911 Legan\'{e}s,
Madrid, Spain}

\author{Francisco Dom\'{\i}nguez-Adame}

\address{Departamento de F\'{\i}sica de Materiales,
Facultad de F\'{\i}sicas, Universidad Complutense, %\\
E-28040 Madrid, Spain}

\maketitle

\begin{abstract}

We introduce a solvable model of a nonlinear double-barrier structure,
described by a generalized effective-mass equation with a nonlinear
coupling term.  This model is interesting in its own right for possible
new applications, as well as to help understand the combined effect of
spatial correlations and nonlinearity on disordered systems.  Although
we specifically discuss the application of the model to electron
transport in semiconductor devices, our results apply to other contexts,
such as nonlinear optical phenomena.  Our model consists of finite width
barriers and nonlinearities are dealt with separately in the barriers
and in the well, both with and without applied electric fields.  We
study a wide range of nonlinearity coupling values.  When the nonlinear
term is only present in the barriers, a sideband is observed in addition
to the main resonance and, as a consequence, the current-voltage
characteristics present two peaks at different voltages.  In this case,
the phenomenology remains basically the same through all the considered
variation of the parameters.  When the well is the component exhibiting
nonlinearity, the results depend strongly on the nonlinear coefficient.
For small values, the current-voltage characteristics exhibit lower and
upper voltage cutoffs.  This phenomenon is not present at higher
nonlinearities, and eventually linear-like behavior is recovered.  We
complete this exhaustive study with an analysis of the effects of having
simultaneously both kinds of nonlinear couplings.  We conclude the paper
with a summary of our results, their technological implications, and a
prospective discussion of the consequences of our work for more complex
systems.

\end{abstract}


\begin{multicols}{2}

\narrowtext

\section{Introduction}

This decade is witnessing a rapid increase of the amount of effort
devoted to the study of one-dimensional (1D) systems, for reasons both
theoretical and technological.  In particular, emphasis is being put on
localization and delocalization in non-spatially-periodic (quasiperiodic
or disordered) problems, which are relevant in very many contexts.  From
the fundamental viewpoint, research is being conducted in two different
main directions.  On the one hand, the generality of localization
phenomena in 1D systems has recently been questioned in a number of
papers, \cite{correlated,JPA} which have shown that the introduction of
short-range correlation in the distribution of inhomogeneities leads to
the formation of bands of extended states.  This unexpected phenomenon
can be understood in terms of the structure of the transmission
coefficient for different segments of the considered system.
\cite{Gennady} On the other hand, nonlinear effects have also been
proposed to counteract the localizing influence of disorder on waves of
any nature (see, e.g., the reviews in Ref.\ \onlinecite{yomismo} and
references therein).  Unfortunately, the results of the interplay
between disorder and nonlinearity depend to a certain extent on the
specific model considered and a general theory of this problem is still
lacking (see, for instance, Refs.~\onlinecite{yomismo} and
\onlinecite{Tsironis1} and references therein).

Following along the above line of research, in this paper we concern
ourselves with the study of a nonlinear double-barrier resonant
structure.  Recently, several authors have reported research on related
devices: For instance, a mean field analysis of multiple resonant
tunneling exhibiting chaotic behavior has been carried out by Presilla,
Jona-Lasinio, and Capasso; \cite{Capasso} in a different context,
directional couplers including nonlinear elements have been investigated
as optical switches.\cite{Tsironis2} The model we propose shares some of
the characteristics of these: Specifically, it is described by a
Schr\"odinger equation where an effective nonlinearity is included and,
being formally close to effective nonlinear Schr\"odinger equations
arising in other problems, it is also very general.  However, we are
going to be mostly interested in its application to resonant tunneling
(RT) through semiconductor heterostructures.  This phenomenon, which
takes place in linear double-barrier structures (DBS), make these
systems very promising candidates for a new generation of ultra-high
speed electronic devices: For instance, a GaAs-Ga$_{1-x}$Al$_x$As DBS
operating at THz frequencies has already been reported in the
literature.\cite{Sollner}

The basic reason for RT to arise in DBS is a quantum phenomenon whose
fundamental characteristics are by now well understood: There exists a
dramatic increase of the electron transmitivity whenever the energy of
the incident electron is close to one of the unoccupied
quasi-bound-states inside the well.\cite{Ricco} In practice, a bias
voltage is applied to shift the energy of this quasi-bound-state of
nonzero width so that its center matches the Fermi level.  Consequently,
the $j-V$ characteristics present negative differential resistance
(NDR).  In actual samples, however, the situation is much more complex
than this simple picture.  This is so even in good-quality
heterostructures, when scattering by dislocations or surface roughness
is negligible.  In particular, {\em inelastic} scattering is always
present in real devices.  Examples of inelastic scattering events are
electron-lattice and electron-electron interactions, in which the energy
of the tunneling electron changes and the phase memory is lost.  Strong
local coupling between electronic and vibrational degrees of freedom
(i.e., when electrons propagate in a deformable charged medium) leads to a
self-induced attractive force which we will account for by means of an
effective nonlinearity as discussed below.  Moreover, Hartree-type
repulsive interactions between electrons can be viewed as a self-induced
repulsive force within an effective medium framework.  There are
currently available very many materials with different optical or
electrical nonlinear properties, and then it may be possible to
purposely build nonlinear DBS structures like those we work with.  In
view of this, the question arises as to whether such nonlinear devices
will have characteristics of interest for applications.  In any case, it
seems that the (intentional or not) introduction of nonlinearity is
going to have nontrivial consequences on the linear phenomenon of RT,
and our purpose here is to study how RT is so modified as well as to
analyze its measurable consequences.

Aside from the above technological and experimental motivations, the
research we summarize in this paper has further goals.  We have recently
studied how short-range spatial correlations affect electron
localization, both theoretically,\cite{JPA,Gennady} (even including
possible three-dimensional effects, see Ref.\ \onlinecite{3D}) and from
the viewpoint of real nanoelectronic devices. \cite{Diez} In this paper,
what we also intend to address is the question of how nonlinearity
modifies our conclusions about linear correlated disordered systems.  We
have already posed this problem in a preliminary work, \cite{PLA} where
we considered concentrated nonlinearities using point-like barriers
described by $\delta$-function potentials.  Nevertheless, such a
situation is rather academic and, in addition, we have shown, for the
case of linear correlated systems, that the results change dramatically
when the width of the barriers is not neglected. \cite{Diez} Therefore,
similar variations may (and should be expected to) occur in the
nonlinear problem when the barrier widths are taken into account, and
this has to be done if any prediction is to be relevant for actual
experiments or devices.

The paper is organized as follows.  In Sec.~II we present our model,
obtained by including a nonlinear coupling in a generalized
effective-mass equation.  We particularize it for a DBS but we insist
that the model is quite general and applicable in different physical
contexts.  We discuss in detail the physics underlying our choice for
the coupling.  We sketch the exact solution of the model, as well as the
way to obtain the transmission coefficient as a function of the
nonlinear couplings and the applied voltage $V$.  For completeness, we
include an appendix containing a brief discussion of the range of
applicability of the equation and its connection with physical
interpretations.  Afterwards, Sec.~III contains the main results and
discussions of our analysis concerning the application to semiconductor
heterostructures, namely electron transmission and $j-V$
characteristics.  We also analyze how the results are changed by the
simultaneous presence of both types of nonlinearity.  Finally, Sec.~IV
concludes the paper with a brief survey of the results and some
prospects on the application of the ideas we have discussed.

\section{Model}

\subsection{Physical grounds and definitions}

As we announced in the introduction, to describe our model we have
chosen to apply it to a specific system: A semiconductor DBS under an
applied electric field.  In the following, we will be using parameters
corresponding to GaAs-Ga$_{1-x}$Al$_x$As as typical values for the
linear DBS, whereas the nonlinear terms will be taken at most as a few
percents of the built-in potential.  Larger nonlinearities are not
considered because they may invalidate from the beginning the
approximations involved in our theoretical calculation and, besides,
they may make impossible to find any material with the desired
properties, at least in the context of electron transport.  Our choices
are thus as follows.  The thickness of the whole structure is $L$ and
the thickness of the well is $d$.  The barriers are assumed to be of the
same thickness (symmetric case) but as will be evident below this is not
a restriction of our approach.  The structure is embedded in a highly
doped material acting as contact, so that the electric field is applied
only in the DBS. We focus on electron states close to the bandgap and
thus we can neglect nonparabolicity effects hereafter.  Then the
one-band effective-mass framework is completely justified to get
accurate results.  For the sake of simplicity, we will further assume
that the electron effective-mass $m^*$ and the dielectric constant are
the same in both materials.  This hypothesis is related to the fact that
we are not interested in high quantitative accuracy, although we note
that the spatial dependence of these parameters can be taken into
account if necessary.

Within this approach, the electron wave function is written as a product
of a band-edge orbital with a slowly varying envelope-function.
Therefore the envelope-function $\psi(z)$ satisfies a generalized
effective-mass equation (we use units such that energies are measured in
effective Rydberg, Ry$^*$, and lengths in effective Bohr radius, a$^*$,
being $1\,$Ry$^*=5.5\,$meV and $1\,$a$^*=100\,$\AA\ in GaAs) given by
\begin{equation}
-\psi_{zz}(z)+\left[V(z)-eFz\right]\,\psi(z)=E\>\psi(z),
\label{Sch}
\end{equation}
where $V(z)$ is the potential term which we discuss below and $F$ is the
electric field applied along the growth direction.  We note, in
connection with the generality of our model, that equations similar to
Eq.~(\ref{Sch}) are used to describe light or other electromagnetic wave
propagation in dielectric superlattices, in an approximation where only
the scalar nature of the waves is taken into account. \cite{Tsironis2}
We now specify our model by choosing what is the potential term $V(z)$.
In order to do that, let us first consider the physics we are trying to
represent with this term.  The DBS can be regarded as an effective
medium which reacts to the presence of the tunneling electron, leading
to a feedback mechanism by which inelastic scattering processes change
the RT characteristics of the device.  It thus follows that $V(z)$ must
contain nonlinear terms if it is to summarize the medium reaction which
comes from the electron-electron and electron-lattice interactions.  The
simplest candidate to contain this feedback process is the charge
density of the electron, which is proportional to $|\psi(z)|^2$.  In our
model, we neglect higher order contributions and postulate that the
potential in Eq.~(\ref{Sch}) has the form
\begin{equation}
\label{Veff}
V(z) = V_0\left\{\left[1+\tilde{\alpha}|\psi(z)|^2\right]\chi_b(z)+
\tilde{\beta}|\psi(z)|^2\chi_w(z)\right\},
\end{equation}
where $V_0$ is the conduction band-offset at the interfaces, and
$\chi_b(z)$ and $\chi_w(z)$ are respectively the characteristic
functions of the barriers and the well,
\begin{mathletters}
\begin{equation}
\chi_b(z)=\left\{\begin{array}{ll} 1, & \mbox{\rm if}\ 0<z<(L-d)/2,\\
                                   1, & \mbox{\rm if}\ (L+d)/2<z<L,\\
                                   0, & \mbox{otherwise},
                 \end{array} \right.
\label{charab}
\end{equation}
\begin{equation}
\chi_w(z)=\left\{\begin{array}{ll} 1, & \mbox{\rm if}\ (L-d)/2<z<(L+d)/2,\\
                                   0, & \mbox{otherwise}.
                 \end{array} \right.
\label{charaw}
\end{equation}
\end{mathletters}
and all the nonlinear physics is contained in the coefficients
$\tilde{\alpha}$ and $\tilde{\beta}$ which we discuss below.

There are two factors that configure the medium response to the
tunneling electron.  First, it goes without saying that there are
repulsive electron-electron Coulomb interactions, which should enter the
effective potential with a positive term proportional to the charge,
i.e., the energy is increased by local charge accumulations, leading to
a positive sign for $\tilde{\alpha}$ and $\tilde{\beta}$.  On the other
hand, in polar semiconductors, the electron polarizes the surrounding
medium creating a local, positive charge density.  Hence the electron
reacts to this polarization and experiences an attractive potential,
which implies $\tilde{\alpha}$ and $\tilde{\beta}$ negative.  This
happens, for instance, in the polaron problem in the weak coupling
limit, which becomes valid in most semiconductors, and where it can be
seen that the lowest band energy state decreases. \cite{Callaway} It is
then clear that in principle any sign would be equally possible for the
coefficients if $\tilde{\alpha}$ and $\tilde{\beta}$ are to represent
the combined action of the polarization of the lattice along with
repulsive electron-electron interactions.  Intuitively, however, it is
most realistic to think that $\tilde{\alpha}$ will be negative, because
a positive nonlinear interaction would arise from negative charge
accumulation in the barriers, which is not likely to occur.  However, as
far as the well is concerned, this is not so, because charge does tend
to accumulate between the two barriers.  This is the case considered,
for instance, in the works of Presilla {\em et al\/.} \cite{Capasso}
where they introduce a term in the Schr\"odinger equation proportional
to the total charge in the well (the integral of the square of the
wavefunction).  Notice as we are going to apply a field to the DBS, high
charge values will not build up between the two barriers, at least for
low doping levels. Therefore, we
will assume that lattice  polarization
effects are stronger than electron-electron interactions inside the
well, thus leading to a negative $\tilde{\beta}$.  On the other hand, we
discuss below mathematical reasons imposing that $\tilde{\alpha}$ and
$\tilde{\beta}$ have to be negative as expected, allowing for it to be
positive only if they are small.  This will be shown there to stem from
the fact that we are studying a boundary value problem for an ordinary
differential equation, and therefore the case of positive couplings is
restricted to a fully dynamical study, which will be the subject of
further work. \cite{moco}

\subsection{Analytical results}

We now work starting from Eq.\ (\ref{Sch}) with the definition in
Eq.~(\ref{Veff}) to cast our equations in a more tractable form.  For
simplicity, and because we are interested in intrinsic DBS features, we
consider that the contacts in which the structure is embedded behave
linearly.  Therefore, the solution of Eq.~(\ref{Sch}) is a linear
combination of traveling waves.  As usual in scattering problems, we
assume an electron incident from the left and define the reflection,
$r$, and transmission, $t$, amplitudes by the relationships
\begin{equation}
\psi(z)=\left\{ \begin{array}{ll} A\left(e^{ik_0z}+re^{-ik_0z}\right)
       & z<0, \\ Ate^{ik_Lz} & z>L,  \end{array} \right.
\label{solution}
\end{equation}
where $k_0^2=E$, $k_L^2=E+eFL$, and $A$ is the incident wave amplitude.
Now we define $\psi(z)=A\phi(z)$, $\alpha=\tilde{\alpha}|A|^2$, and
$\beta=\tilde{\beta}|A|^2$.  Notice that $\alpha$ and $\beta$ are
dimensionless parameters and that they depend on the amplitude of the
incoming wave, which will be relevant later. Using Eq.~(\ref{Sch}) we get
\begin{equation}
-\phi_{zz}(z)+[V(z)-eFz-E]\,\phi(z)=0.
\label{fi}
\end{equation}

To solve the scattering problem in the DBS  we develop a
similar approach to that given in Ref.~\onlinecite{Knapp}.  Since
$\phi(z)$ is a complex function, we write $\phi(z) = q(z) \exp[
i\gamma(z)]$, where $q(z)$ and $\gamma(z)$ are real functions.
Inserting this factorization in Eq.~(\ref{fi}) we have $\gamma_z (z) =
q^{-2}(z)$ and
\begin{eqnarray}
-q_{zz}(z)+{1\over q^3(z)}+\left[V_0\chi_b(z)-eFz-E\right]\,q(z) & + &
\nonumber \\
+V_0\left[\alpha\chi_b(z)+\beta \chi_w(z)\right]q^3(z) & = & 0.
\label{q}
\end{eqnarray}
This nonlinear differential equation must be supplemented by appropriate
boundary conditions. However, using Eq.~(\ref{solution}) this problem
can be converted into a initial conditions equation. In fact, it is
straightforward to prove that
\begin{equation}
\label{ic}
q(L)=k_L^{-1/2},\>q_z(L)=0,
\end{equation}
and that the transmission coefficient is given by
\begin{equation}
\tau=\,{4k_0q^2(0)\over 1+2k_0q^2(0)+k_0^2
q^4(0)+q^2(0)q_z^2(0)}.
\label{tau}
\end{equation}
Hence, we can integrate numerically (\ref{q}) with initial conditions
(\ref{ic}) backwards, from $z=L$ up to $z=0$, to obtain $q(0)$ and
$q_z(0)$, thus computing the transmission coefficient for given
nonlinear couplings $\alpha$ and $\beta$, incoming energy $E$ and
applied voltage $V=FL$.

Once the transmission coefficient has been computed, and recalling that
contacts are linear media, the tunneling current density at a given
temperature $T$ for the DBS can be calculated within the
stationary-state model from
\begin{mathletters}
\label{eq2}
\begin{equation}
j(V)={m^*ek_BT\over 2\pi^2\hbar^3}\,\int_0^\infty\> \tau(E,V)N(E,V)\,dE,
\label{eq2a}
\end{equation}
where $N(E,V)$ accounts for the occupation of states to both sides of
the device, according to the Fermi distribution function, and it is
given by
\begin{equation}
N(E,V)=\ln\left(\frac{1+\exp[(E_F-E)/k_BT]}{1+\exp[(E_F-E-eV)/k_BT]}
\right),
\label{eq2b}
\end{equation}
\end{mathletters}
where $k_B$ is the Boltzmann constant.

\section{Results and discussions}

In our calculations we have considered a GaAs-Ga$_{0.65}$Al$_{0.35}$As
double-barrier structure with $L=3d=150\,$\AA.  The conduction-band
offset is $V_0=250\,$meV.  In the absence of applied electric field and
nonlinearities, there exist a single, very narrow resonance with
$\tau\sim 1$ below the top of the barrier, with an energy of
$80.7\,$meV, and hence the well supports a single quasi-bound state.
When voltage is applied, the energy of the quasi-bound state level is
lowered and a strong enhancement of the current arises whenever the
Fermi level matches this resonance, thus leading to the well-known RT
phenomenon.  As we will see, it is enough to have a small amount of
nonlinearity to have this picture changed dramatically.  In this
respect, before entering our report, we want to stress that the
coefficients $\alpha$ and $\beta$ that we have defined depend not only
on the intrinsic characteristics of the material but also on the energy
of the incoming electrons, as they both include the factor $|A|^2$.
Therefore, it makes sense to study different values for the two
coefficients, because even for the same device transport properties can
be very different for different incoming electrons.
\begin{figure}
\setlength{\epsfxsize}{5.2cm}
\centerline{\mbox{\epsffile{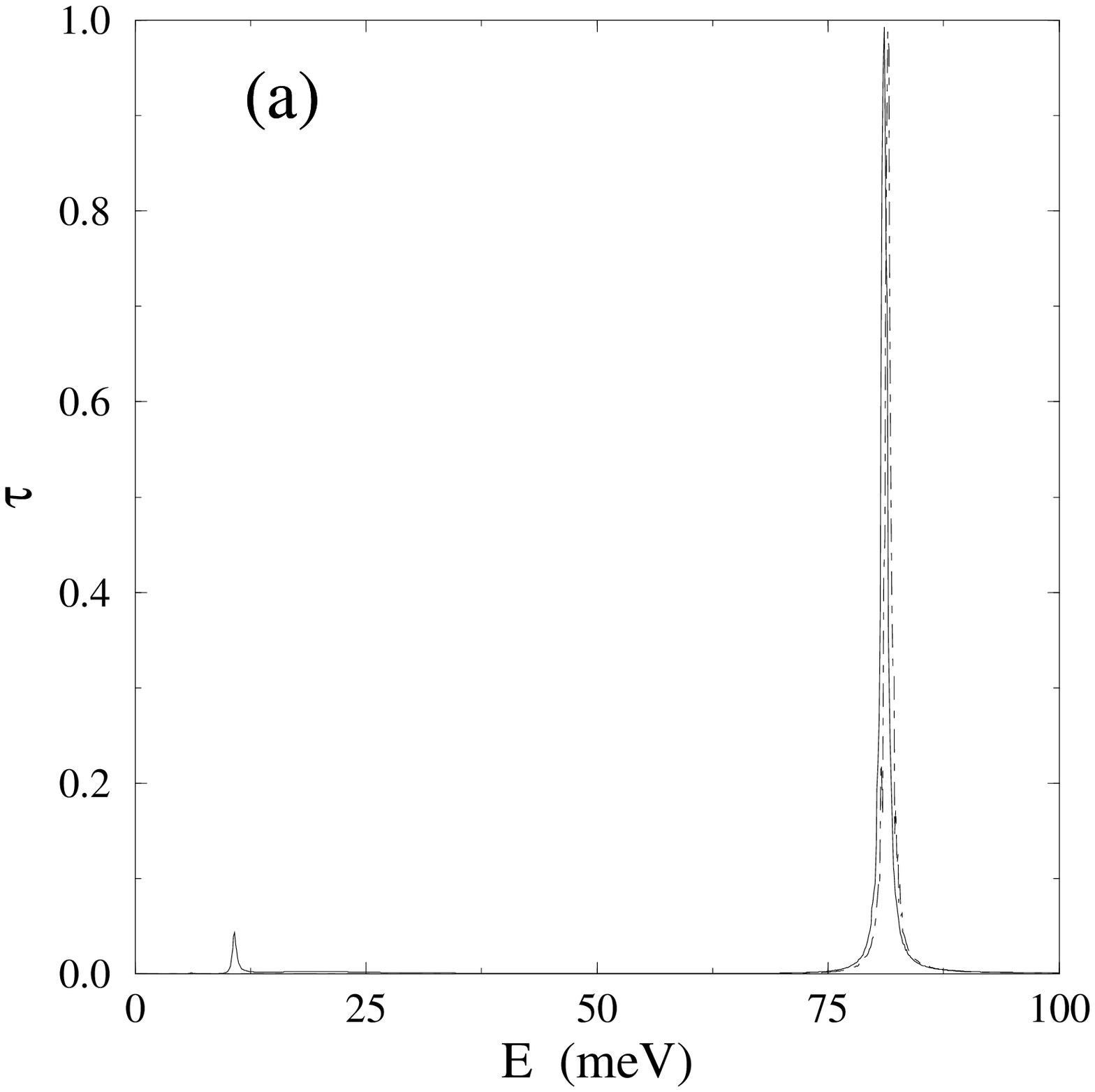}}}
\setlength{\epsfxsize}{5.2cm}
\centerline{\mbox{\epsffile{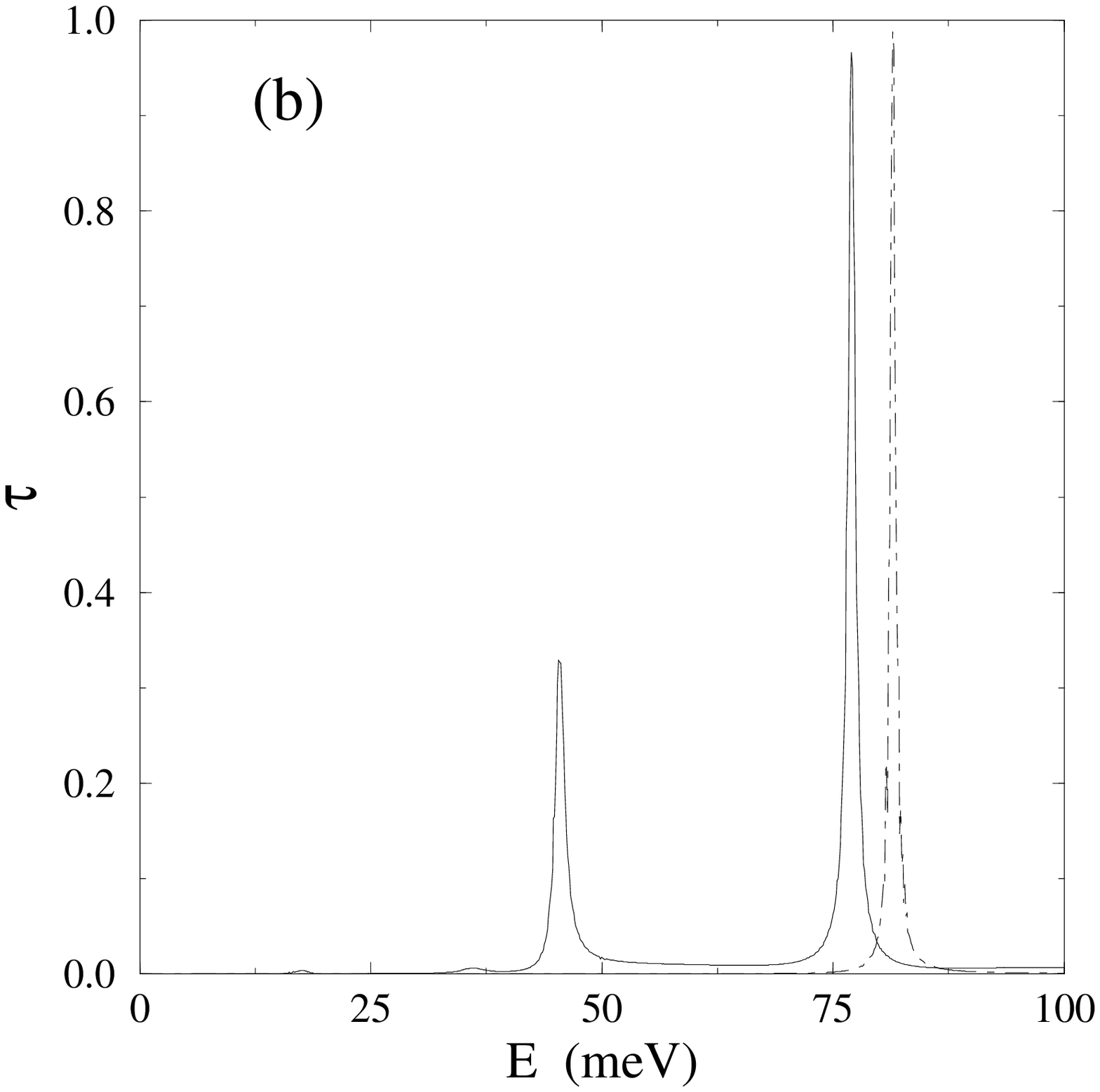}}}
\setlength{\epsfxsize}{5.2cm}
\centerline{\mbox{\epsffile{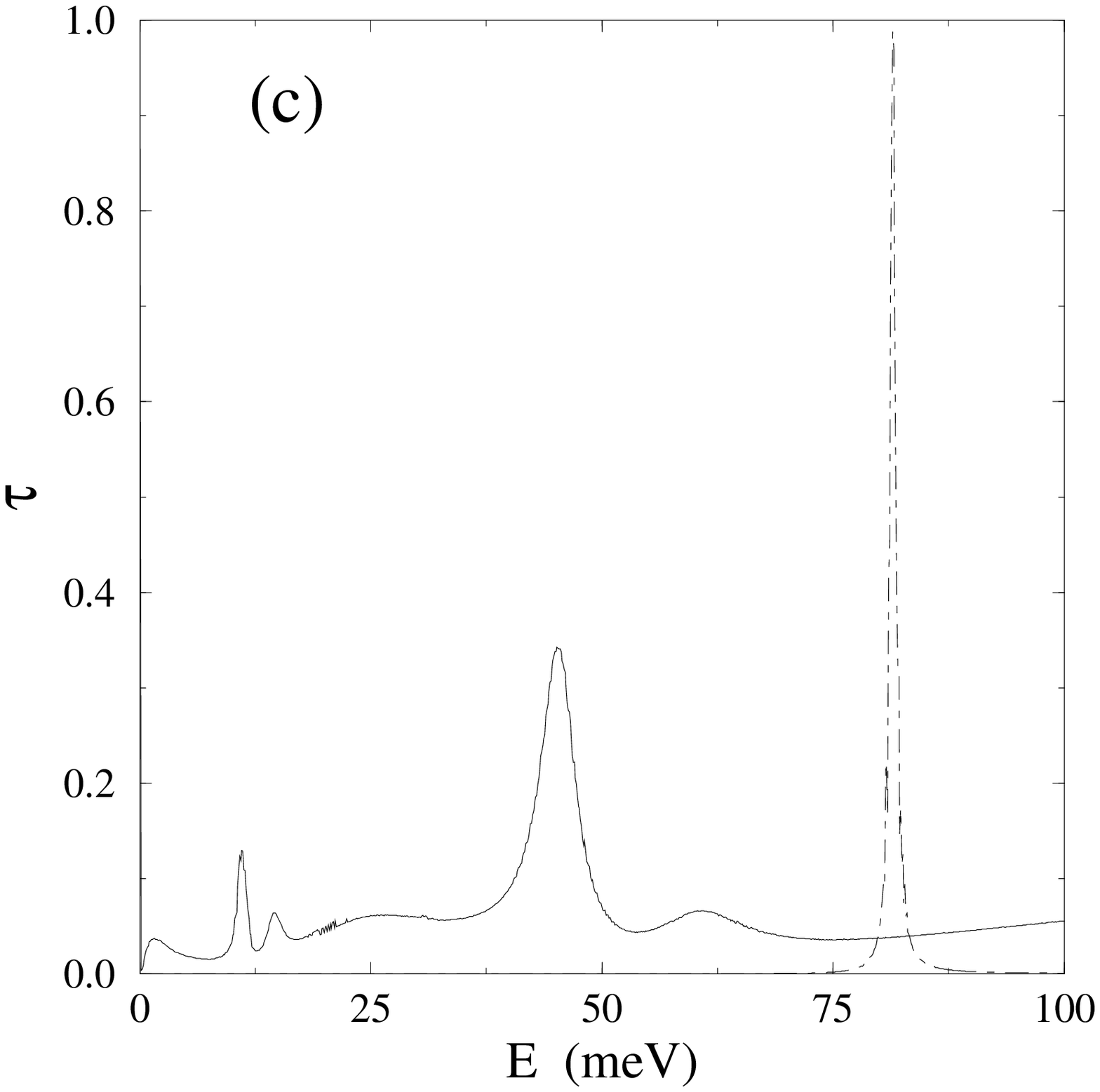}}}
\caption{Transmission coefficient $\tau$ as a function of the
electron energy at zero bias for (a) $\alpha=-0.001$, (b) $-0.01$, and (c)
$-0.1$.
For comparison, dashed line indicates the result for $\alpha=0$.}
\label{fig1}
\end{figure}

\subsection{Nonlinear barriers}

% Parte de alpha solo

We will begin by discussing the effects of having nonlinearity in the
barriers only ($\beta=0$), and our starting point will be to report on
zero field transmission properties.  We will consider three values for
$\alpha$, $\alpha=-0.001, -0.01,\,-0.1$, as representatives of very
different orders of magnitude.  Hereafter, for brevity of language we
will call {\em large} values of the nonlinearity to those large in
absolute value.  For the purposes of the present paper we will regard
values larger than those quoted above as not physically realizable,
although in principle materials with such exotic properties might be
found; smaller values behave mostly like the linear system.
Figure~\ref{fig1} shows the transmission coefficient as a function of
the incoming energy for these three different values of the coefficient
$\alpha$, when the DBS is placed under zero bias.  It is seen that for
the two lower values there exist two peaks, but the smallest one is only
well differentiated for $\alpha=-0.01$.  The main peak is centered at
$\sim 75\,$meV (close to the linear case peak) and the sideband is
centered at $\sim 40\,$meV for $\alpha=-0.01$.  Compared to the purely linear
case, the
main peak not only shifts to lower energies but also broadens, in a
similar fashion to what happens in RT when inelastic effects arise, and
this shifting and broadening is larger when $\alpha$ increases.  Upon
further increasing $\alpha$, the sideband moves below zero, whereas the
main peak reduces appreciably, as shown in Fig.~\ref{fig1}(c).  We thus
see that the effect of strong nonlinearity becomes simply to move down
and broaden the main peak, and therefore the existence of two peaks
belongs only to a limited range of $\alpha$.

\begin{figure}
\setlength{\epsfxsize}{8.0cm}
\centerline{\mbox{\epsffile{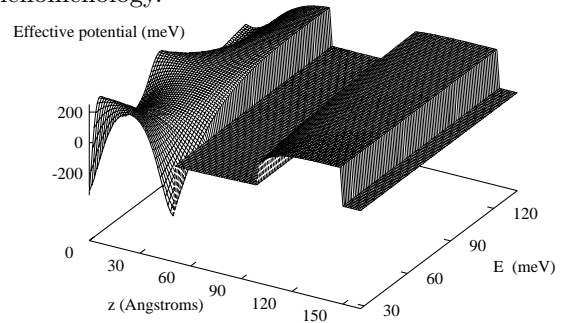}}}
\caption{Effective potential $V_{ef\!f}$ as a function of $z$ and the
incoming electron energy $E$ at zero bias, for $\alpha=-0.01$.}
\label{fig2}
\end{figure}

In order to gain insight into this phenomenon, we can rewrite
Eq.~(\ref{fi}) as follows
\begin{mathletters}
\label{defi}
\begin{equation}
-\phi_{zz}(z)+V_{ef\!f}(z,E)\,\phi(z)=E\>\phi(z),
\label{defia}
\end{equation}
where we have defined an effective potential as follows
\begin{equation}
V_{ef\!f}(z,E)=V_0\chi_b(z)[1+\alpha q^2(z)]-eFz.
\label{defib}
\end{equation}
\end{mathletters}
Thus (\ref{defi}) is a Schr\"odinger-like equation for an effective
potential due to nonlinearity plus the linear potential and the built-in
potential of the DBS. This effective potential depends not only on $z$
but also on the incoming energy $E$ through the function $q(z)$.  Let us
analyze its meaning in the zero field situation.  Since the envelope
function changes under RT conditions, and also does $q(z)$, it should be
expected that $V_{ef\!f}(z,E)$ undergoes severe variations whenever $E$
is close to one of the RT peaks.  This is indeed the case, as shown in
Fig.~\ref{fig2}, where $V_{ef\!f}(z,E)$ is displayed at zero bias for
$\alpha=-0.01$ as a representative example of all the studied values.
Notice that nonlinear effects have negligible effects on the shape of
the effective potential in the right barrier, besides a slight band
bending at the interface $z=(L+d)/2$ at low energies.  However, the
potential in the left barrier region differs significantly from the
original square-barrier shape.  Out of energy resonances, the effective
barrier height at $z=0$ is lower than $V_0$, whereas at resonances it
takes the value $\sim V_0$.  This is clearly related to the ability of
the electrons to go through the structure for the resonant energies and
their subsequent low probability of remaining inside the barrier.  Hence
the effective potential presents two local maxima in the plane $z=0$ as
a function of the incoming energy $E$, matching the values of the main
resonance and the sideband above discussed.  From Fig.~\ref{fig2} it is
clear that at the energy of the main resonance the effective potential
is quite similar to the built-in DBS potential, just producing a small
shift of the quasi-bound-state in comparison to the linear RT process.
Concerning the sideband, the effective potential presents a deep minimum
at the interface $z=(L-d)/2$, which originates another
quasi-bound-state, thus explaining the origin of this lower RT peak.
Indeed, the fact that this added well is responsible for the peak is
confirmed by observations of the effective potential when fields are
applied, as discussed below.  Another reason supporting this hypothesis
is that for the smallest $\alpha$ the sideband is located at a lower
energy, which makes sense since in that case the depth of the extra well
is smaller.  We thus provide a complete coherent picture of the zero
field tunneling phenomenology.

\begin{figure}
\setlength{\epsfxsize}{5.0cm}
\centerline{\mbox{\epsffile{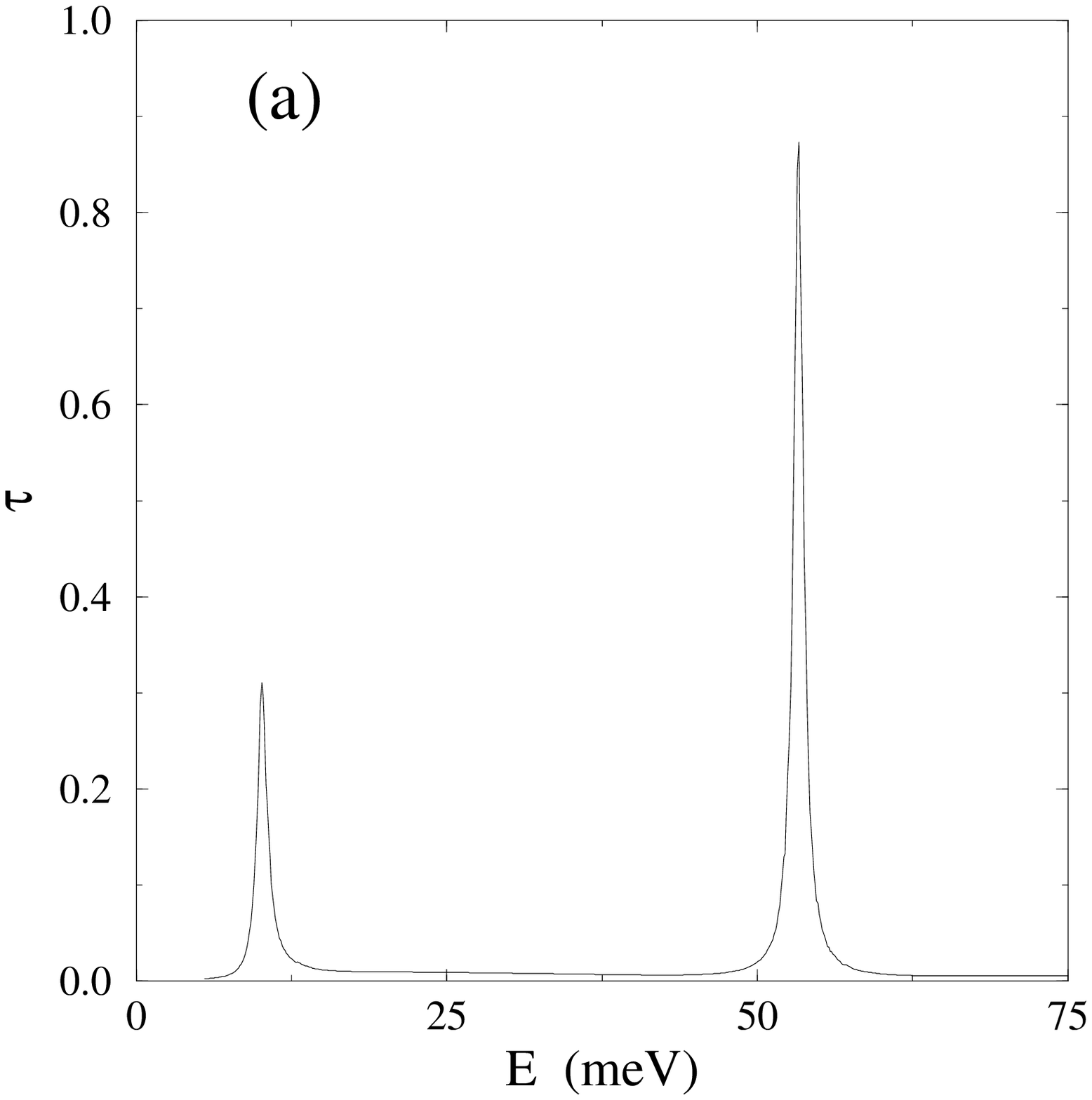}}}
\setlength{\epsfxsize}{5.0cm}
\centerline{\mbox{\epsffile{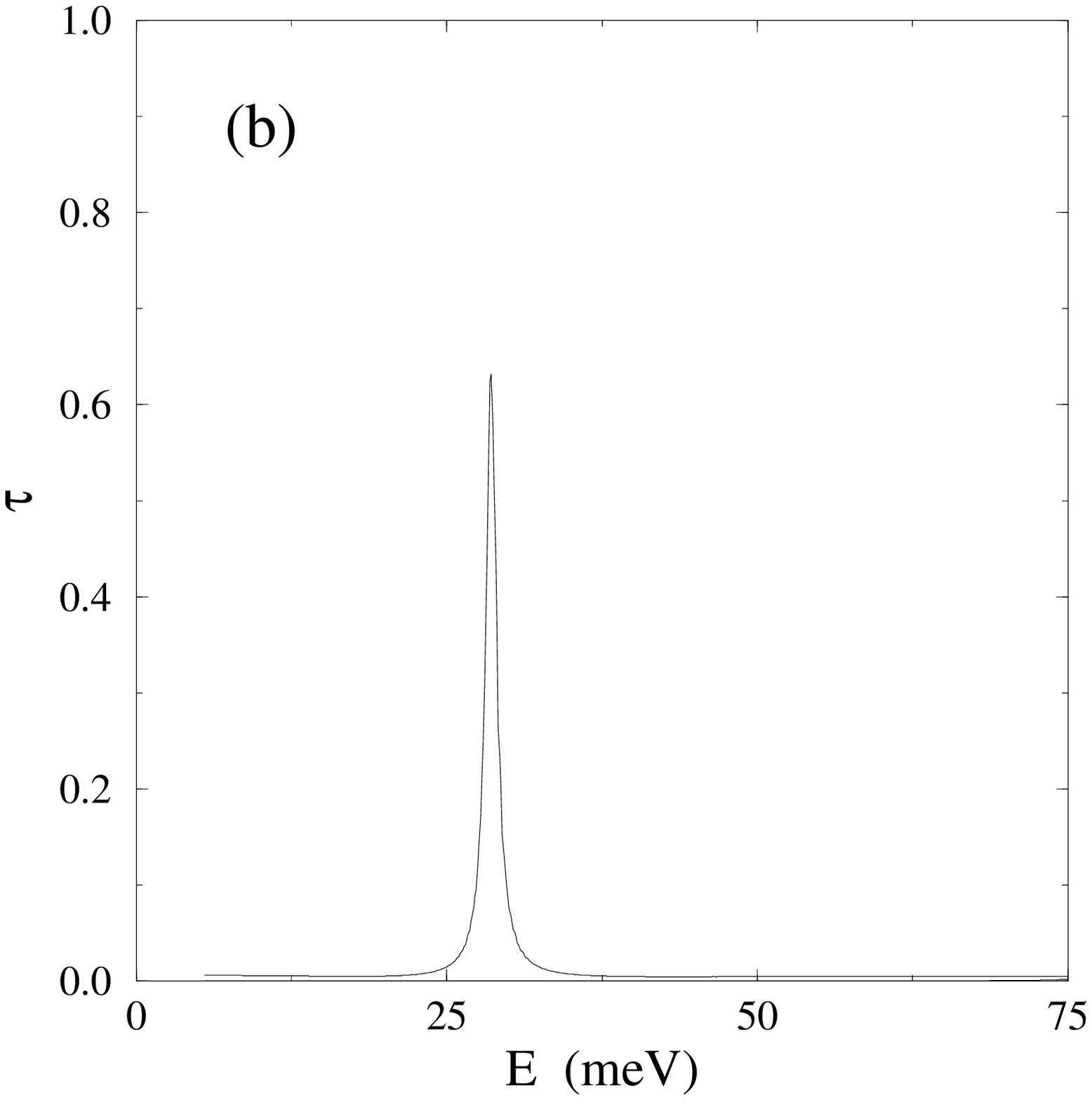}}}
\caption{Transmission coefficient $\tau$ as a function of the
electron energy for $\alpha=-0.01$ with (a) $V=0.05\,$volts and (b)
$0.10\,$volts.} \label{fig1bis}
\end{figure}

\begin{figure}
\setlength{\epsfxsize}{3.8cm}
\centerline{\mbox{\epsffile{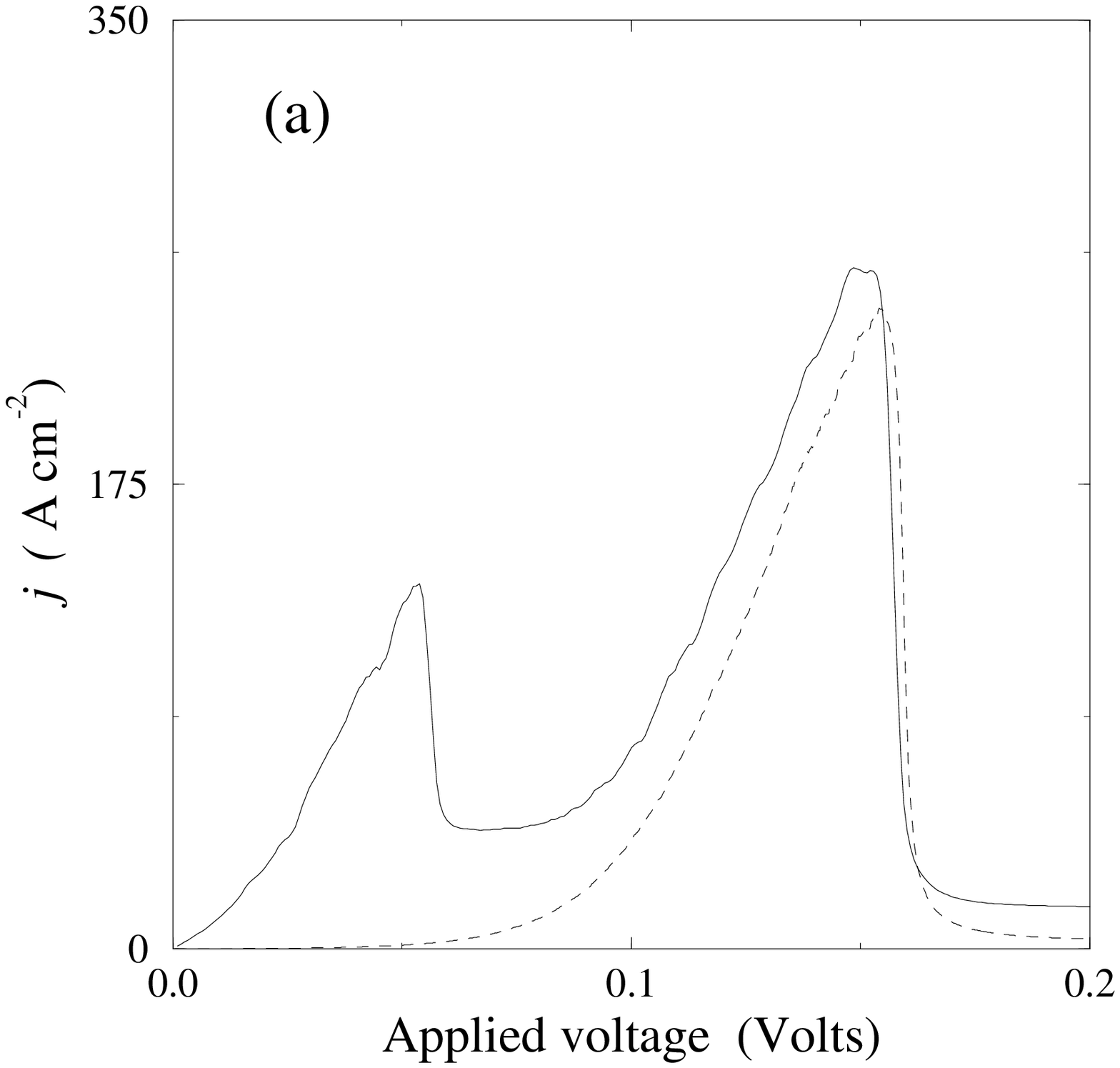}}}
\setlength{\epsfxsize}{3.8cm}
\centerline{\mbox{\epsffile{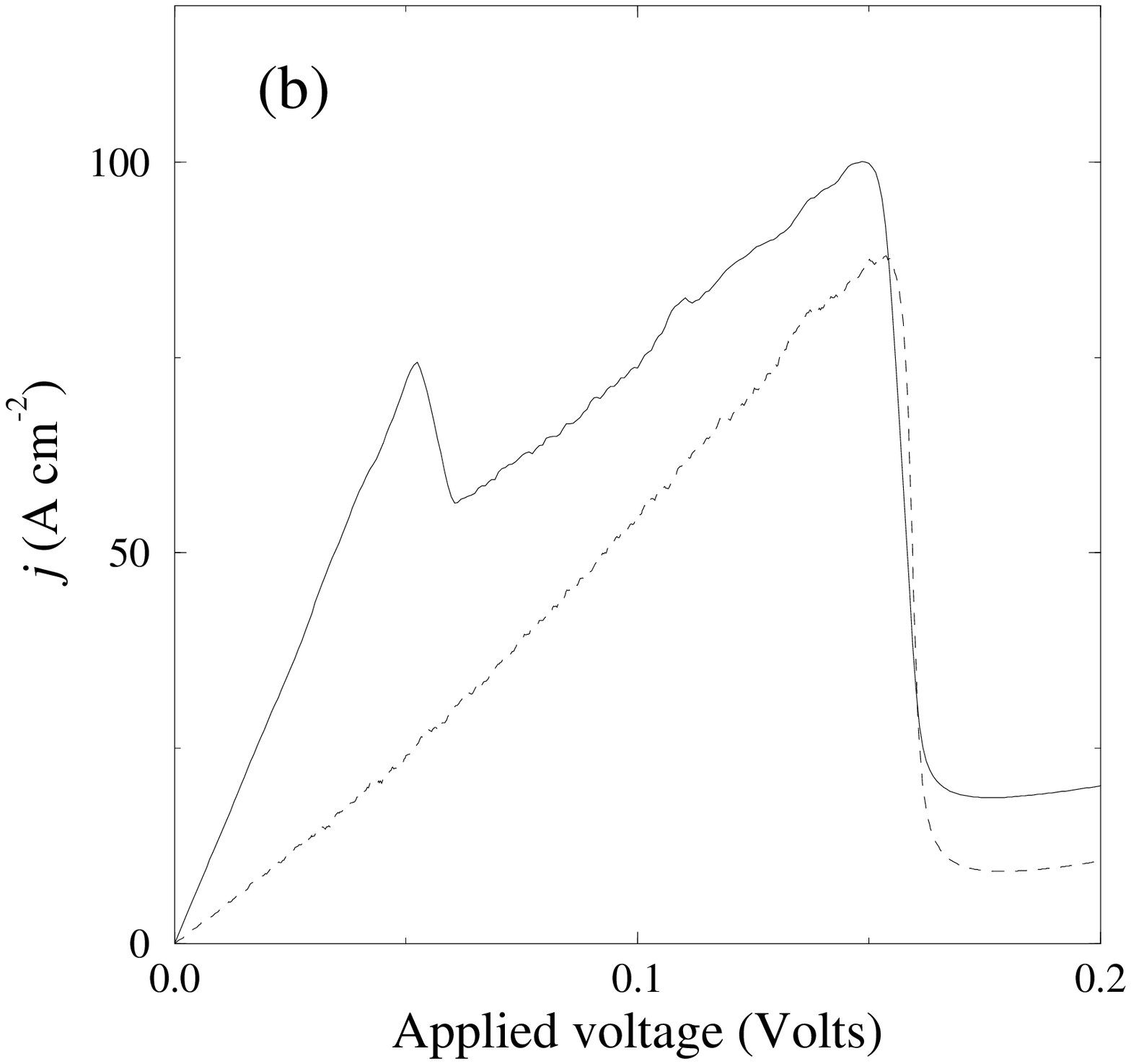}}}
\setlength{\epsfxsize}{3.8cm}
\centerline{\mbox{\epsffile{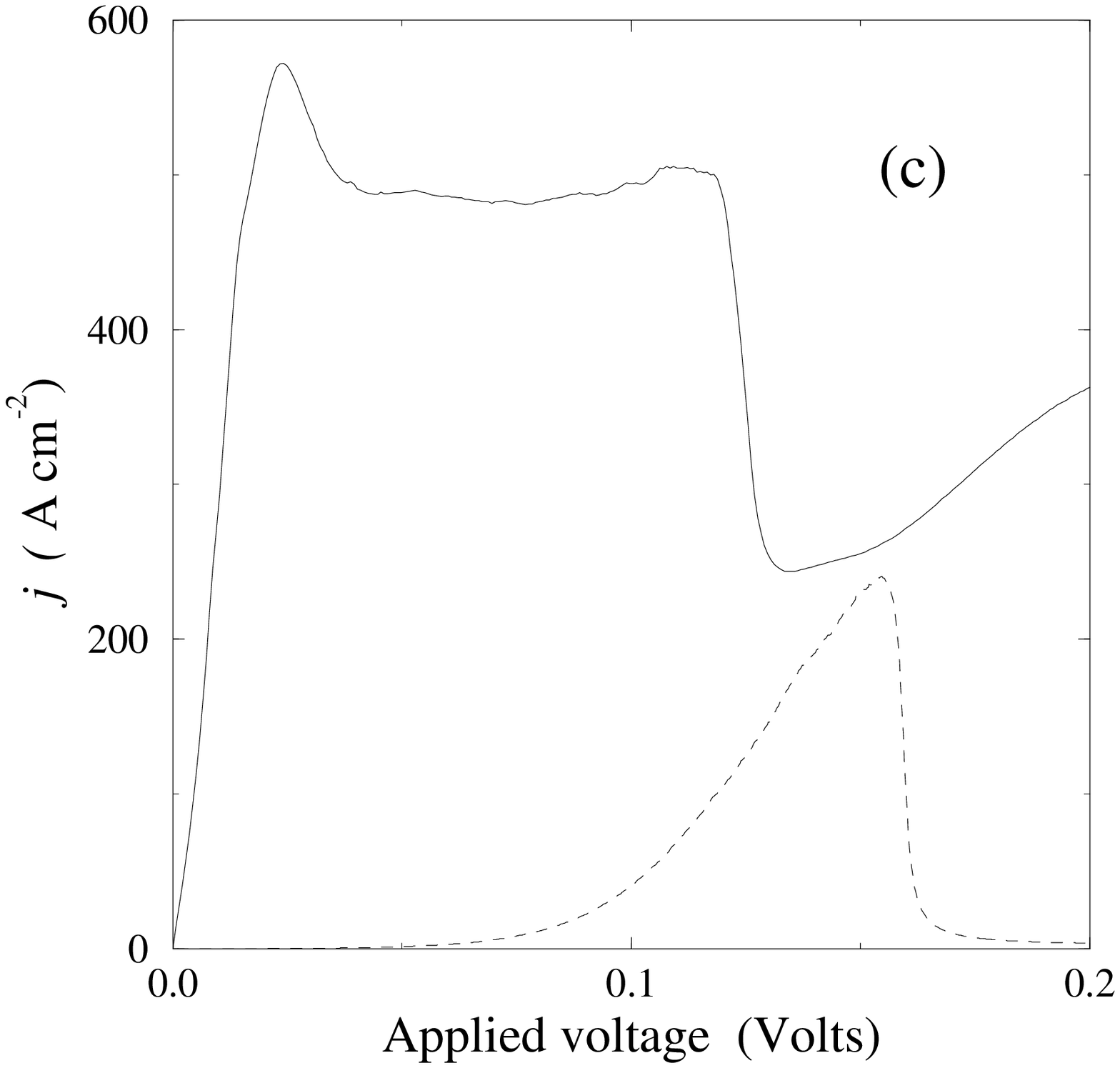}}}
\setlength{\epsfxsize}{3.8cm}
\centerline{\mbox{\epsffile{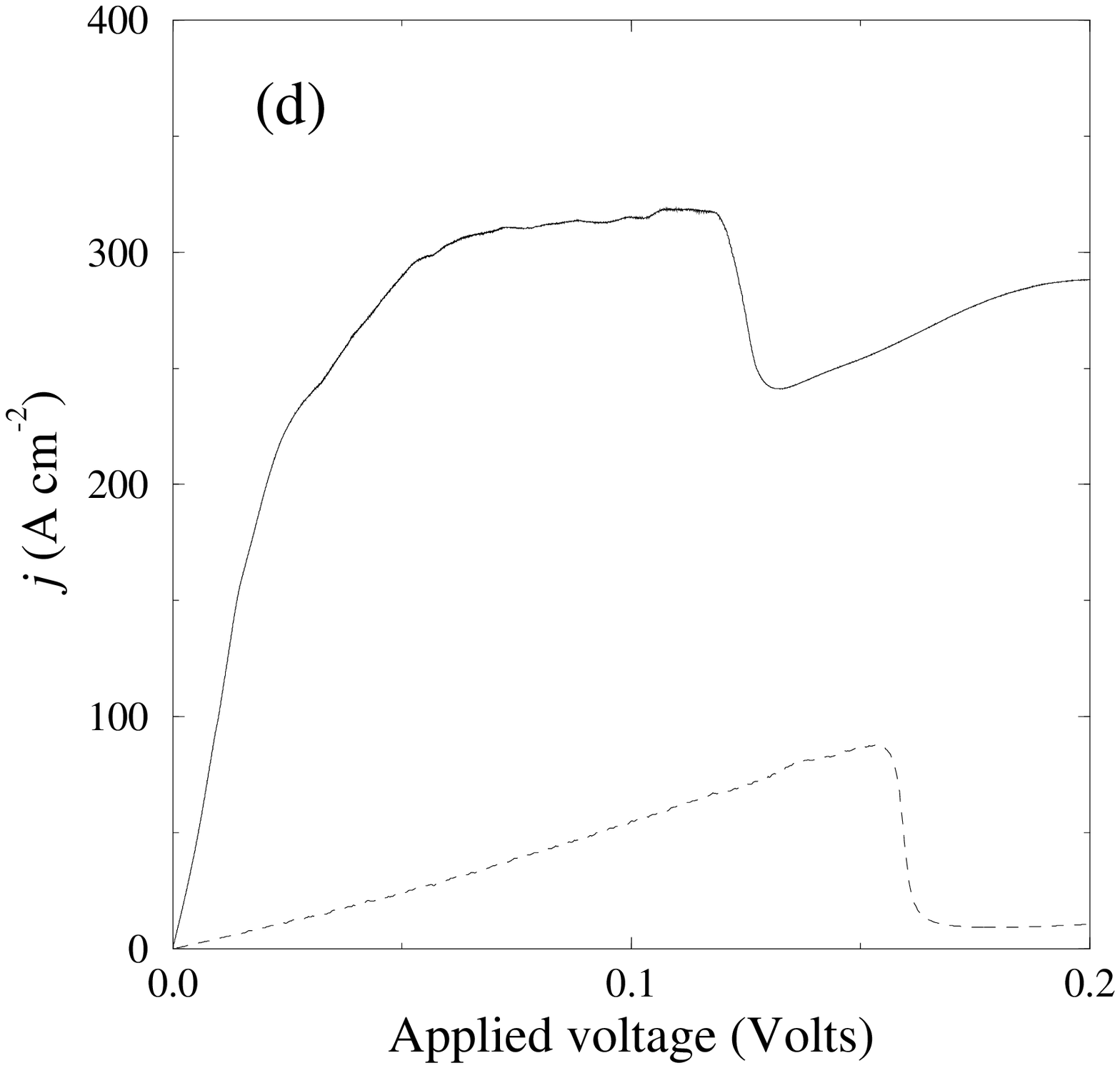}}}
\caption{Computed $j-V$ characteristics for $E_F=27.5\,$meV,
$\alpha=-0.01$  at (a) $T=77\,$K and (b) room temperature,
and $\alpha=-0.1$  at (c) $T=77\,$K and (d) room temperature.
For comparison, dashed line indicates the result for $\alpha=0$.}
\label{fig3}
\end{figure}

We now consider the effect of a bias imposed on the DBS and show an
example of the general behavior for all the considered values of
$\alpha$ in Fig.~\ref{fig1bis}.  The transmission peaks are shifted to
smaller values of energy, leading first, in the cases when there is a
sideband, to the suppression of this subsidiary peak, as can be seen by
comparing Figs.\ \ref{fig1bis} (a) and (b).  In this respect, we have to
say that such shift to lower energies is similar to that found in linear
(coherent) RT. Now we can conclude the interpretation of our findings in
terms of the effective potential as announced.  By reasoning in the same
way as above but taking the field into account, we have concluded that
the fact that the resonance responsible for the peak shifts towards
$E=0$ is due to the influence of the applied field, which increases the
depth of the secondary well; the second resonance goes subsequently
below zero and ceases to be seen in the transmission coefficient.
Afterwards, for higher fields, even the main peak is suppressed by the
same mechanism, although if even higher fields are applied, new levels
are bound by such a deeper well.  We do not go further into the details
of those as we are interested in the small fields suitable for our
approximations and for application to actual devices.
We are now in a position where we are able to comment on the $j-V$
characteristics, computed from (\ref{eq2}).  We have set two different
temperatures ($77\,$K and room temperature) and compared these curves
with those obtained in linear RT. The Fermi energy was $E_F=27.5\,$meV.
We discuss the lower $\alpha$ cases first, shown in Fig.~\ref{fig3}. We have
not plotted the curve for $\alpha=-0.001$ as it turns out to be almost
indistinguishable from the linear one, which always shows a single NDR
peak.  When $\alpha=-0.01$, at T$=77\,$K nonlinearity causes the
occurrence of a second peak at a lower voltage, clearly related to the
sideband in the transmission coefficient.
On increasing temperature up
to room values, both peaks merge into a single one because of the
broadening of the Fermi-Dirac distribution function.  It is important to
notice that the $j-V$ characteristics collect all the main features
found in the transmission coefficient, namely a downwards shifting of
the main RT peak, its corresponding broadening due to inelastic effects
and the appearance of the sideband, as shown in Fig.~\ref{fig3}(a) and
(b) as a smaller NDR signature at $V=0.05\,$volts.  As for the results
for $\alpha=-0.1$, they exhibit related features although somewhat less
prominent and shifted towards lower voltages [see Fig~\ref{fig3}(c) and (d)].
The disappearance of the
sideband for this nonlinearity value leads to a structure of the $j-V$
curve different than the previous one, which consists of a plateau
starting at low voltage values.  This is connected to the nonzero value
of the transmission coefficient [see Fig.\ \ref{fig1}(c)] for all
energies, which allows for transmission even in non-RT conditions.  The
end of the plateau is an NDR interval which is associated to the broad
peak originated by the linear resonance.  In addition, the current is
twice as large as for the previous case.  Following the plateau, the
current drops down but it does not reach zero values, rather it keeps at
a value about half that of the plateau.  This is even more clear from
the room temperature plot, which shows practically no features, i.e.,
loosely speaking, the current tends to be constant.  The current values
the DBS supports are clearly larger than in the linear and weakly
nonlinear cases.

\subsection{Nonlinear well}

\begin{figure}
\setlength{\epsfxsize}{4.8cm}
\centerline{\mbox{\epsffile{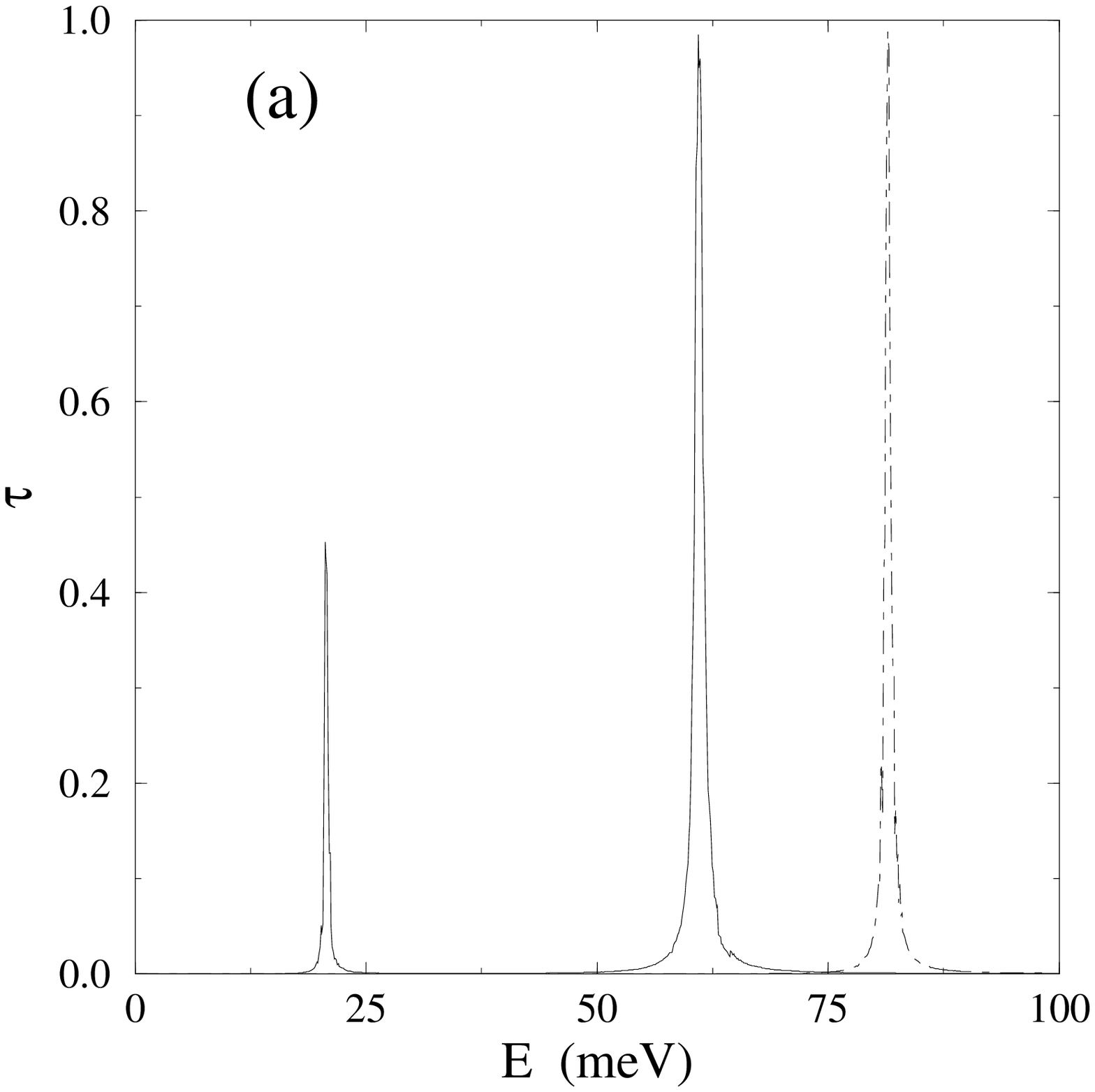}}}
\setlength{\epsfxsize}{4.8cm}
\centerline{\mbox{\epsffile{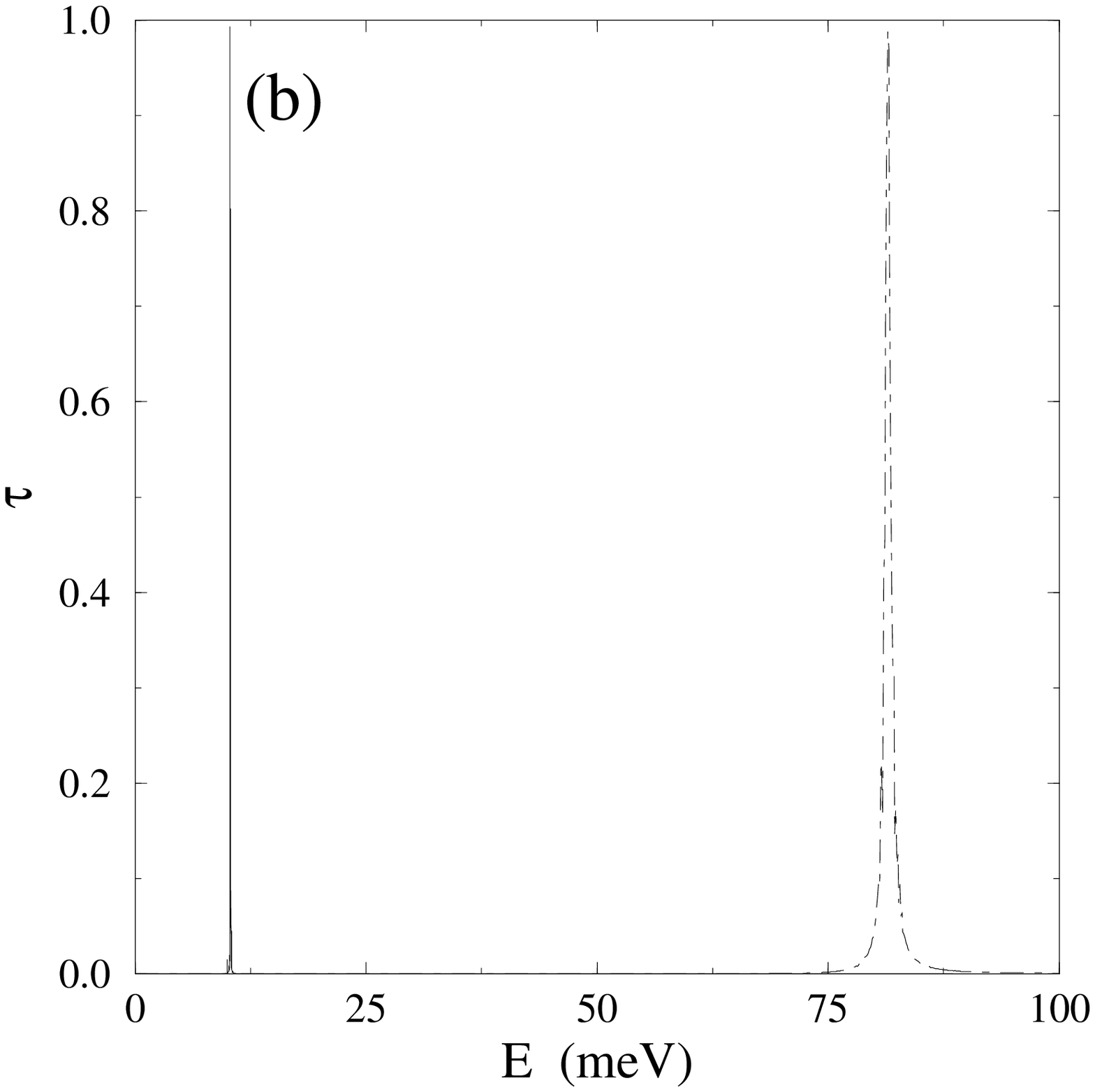}}}
\setlength{\epsfxsize}{4.8cm}
\centerline{\mbox{\epsffile{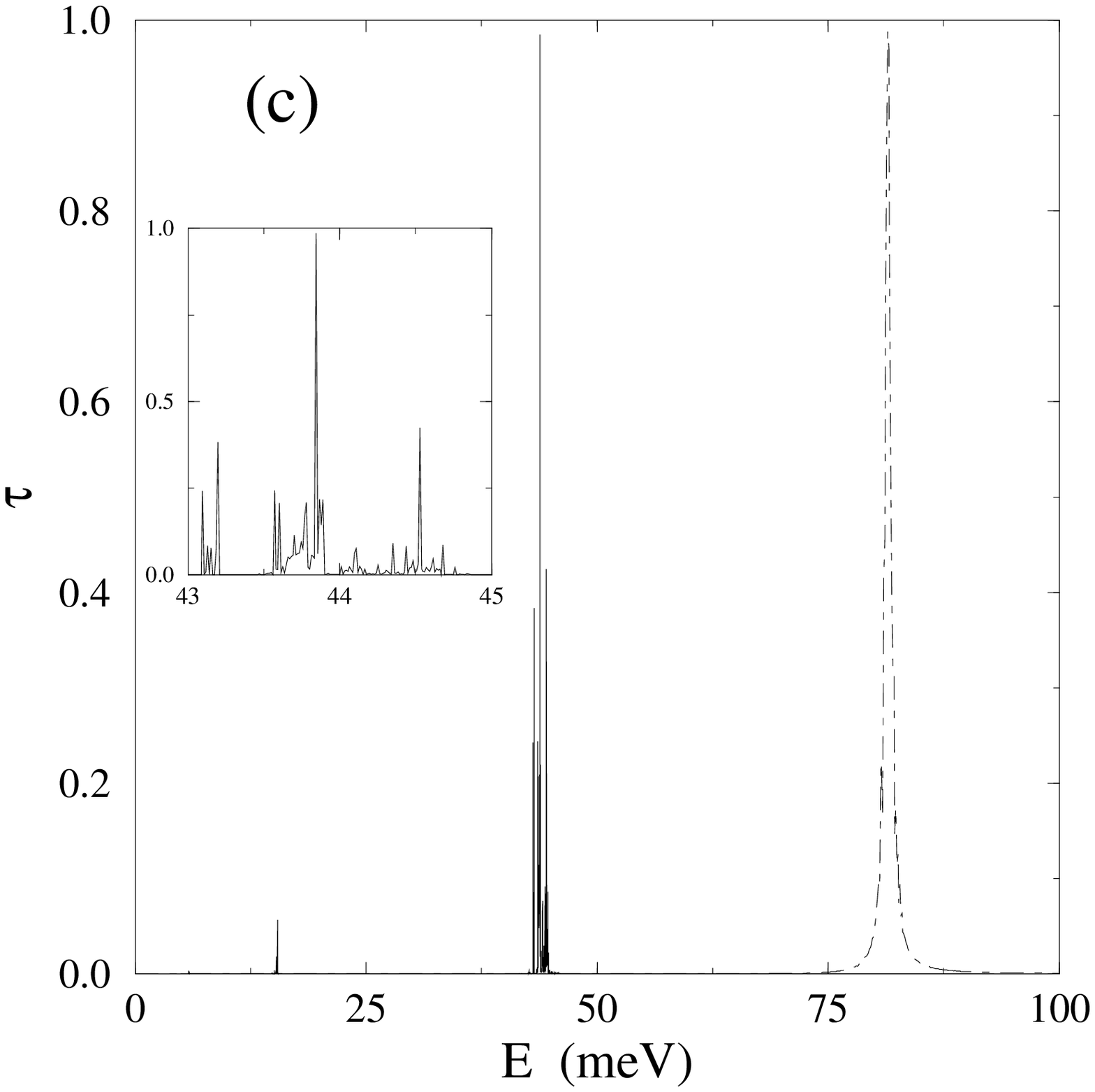}}}
\caption{Transmission coefficient $\tau$ as a function of the
electron energy for (a) $\beta=-0.001$, (b) $-0.01$, and (c) $-0.1$.
The inset shows an enlarged view of the transmission coefficient close to
$45\,$meV.
For comparison, dashed line indicates the result for $\beta=0$.}
\label{fig4}
\end{figure}

\begin{figure}
\setlength{\epsfxsize}{9.0cm}
\centerline{\mbox{\epsffile{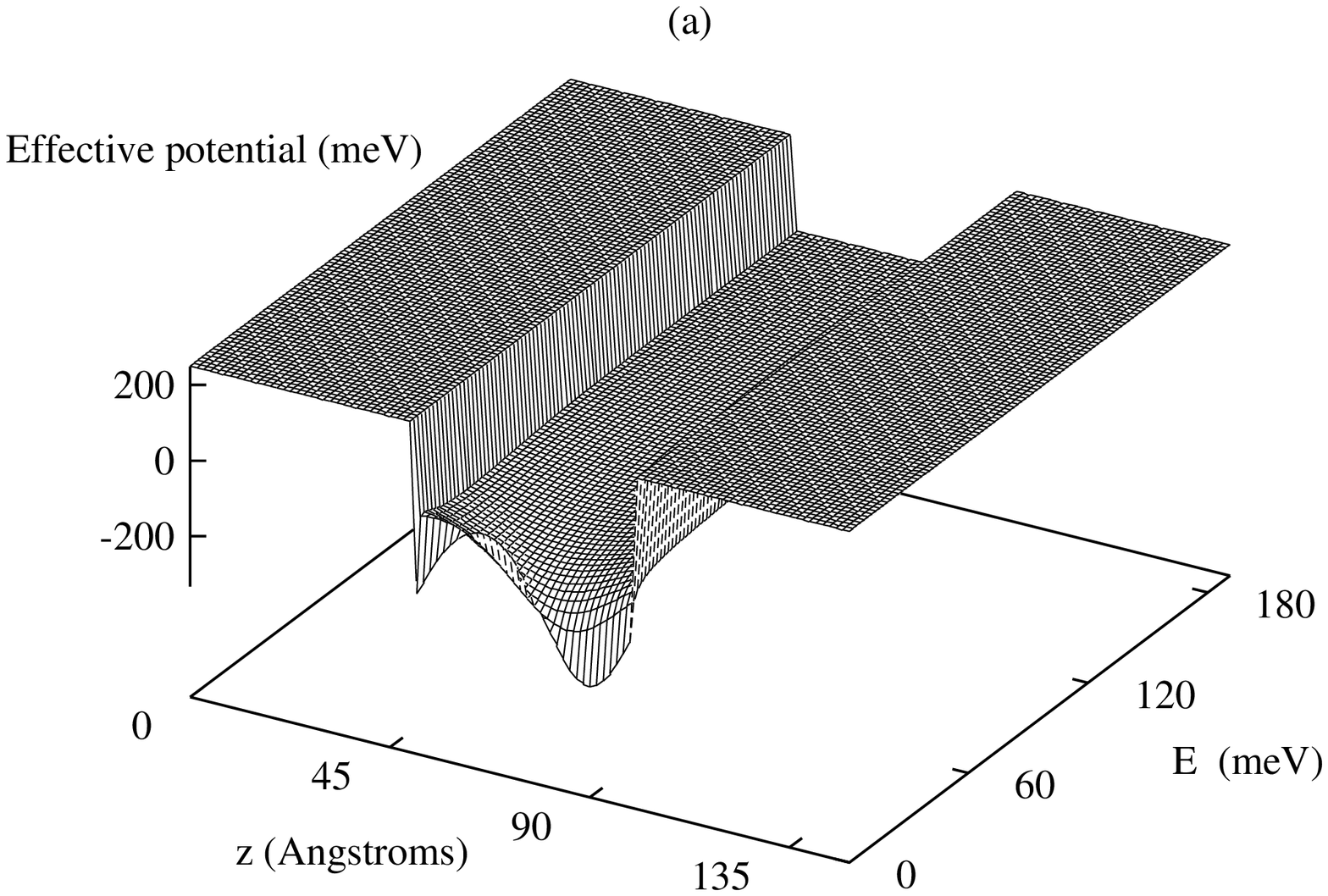}}}
\setlength{\epsfxsize}{9.0cm}
\centerline{\mbox{\epsffile{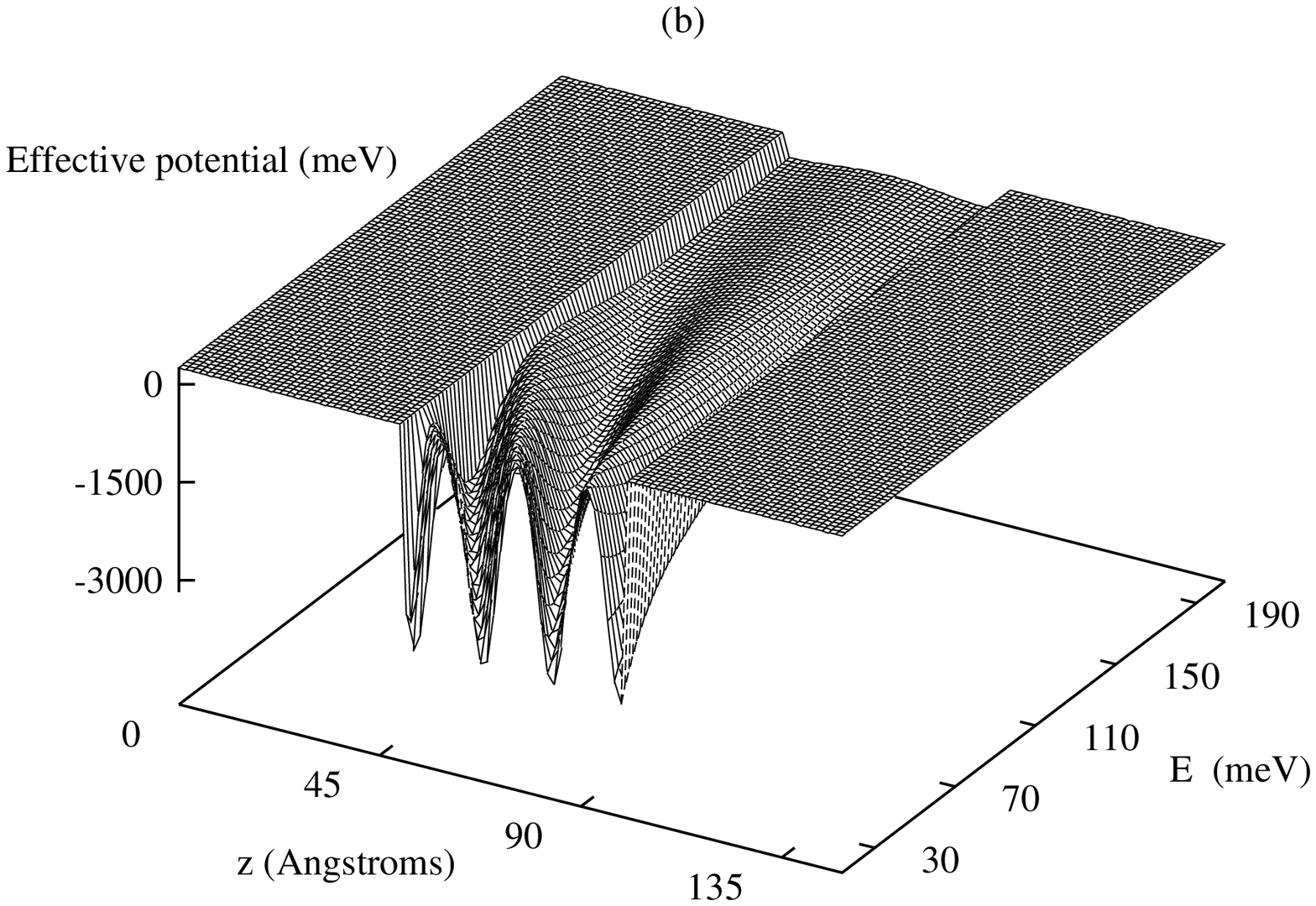}}}
\caption{Effective potential $V_{ef\!f}$ as a function of $z$ and the
incoming electron energy $E$ at zero bias for (a) $\beta=-0.001$ and
(b) $-0.1$.}
\label{fig5}
\end{figure}

\begin{figure}
\setlength{\epsfxsize}{3.8cm}
\centerline{\mbox{\epsffile{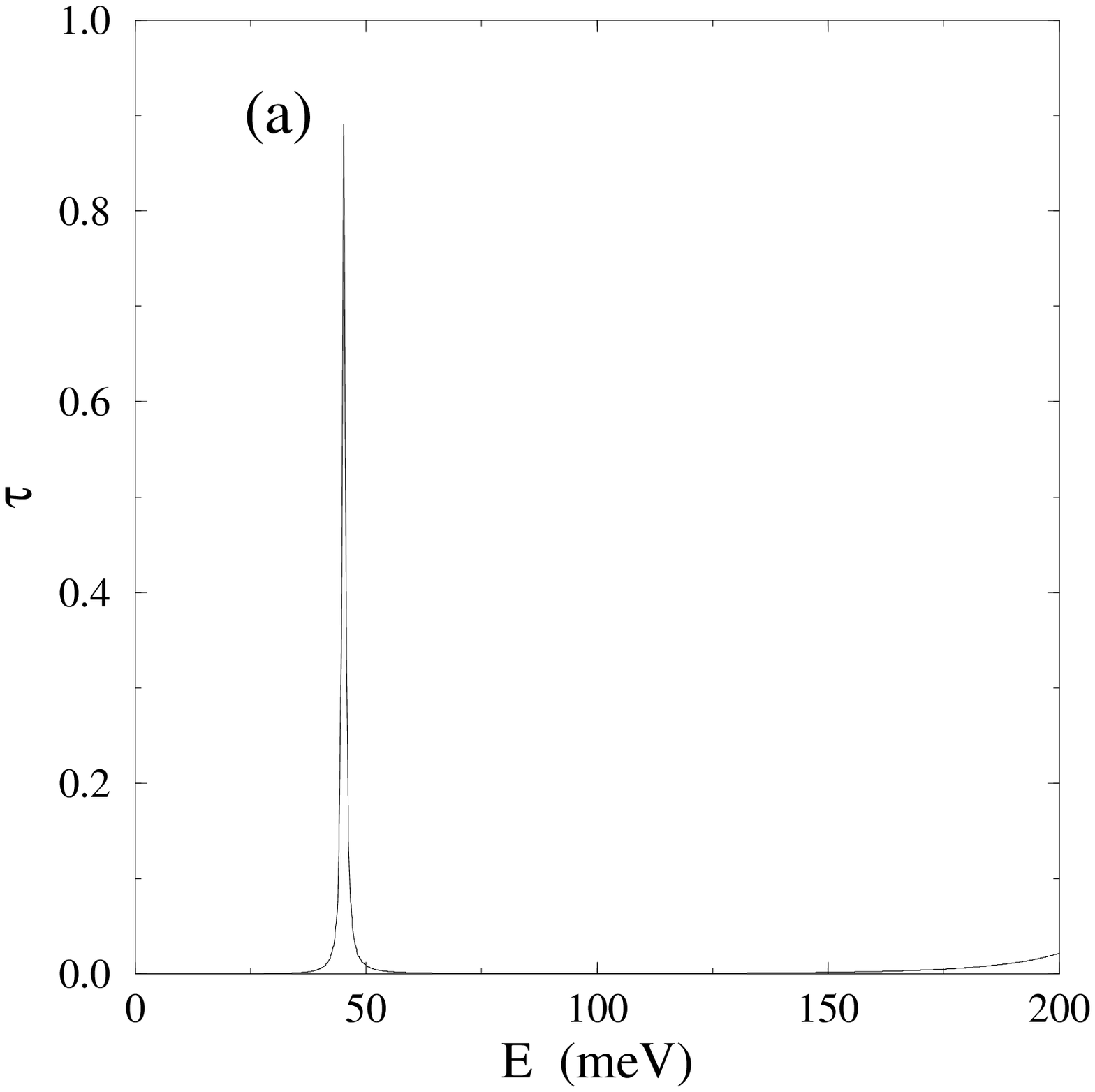}}}
\setlength{\epsfxsize}{3.8cm}
\centerline{\mbox{\epsffile{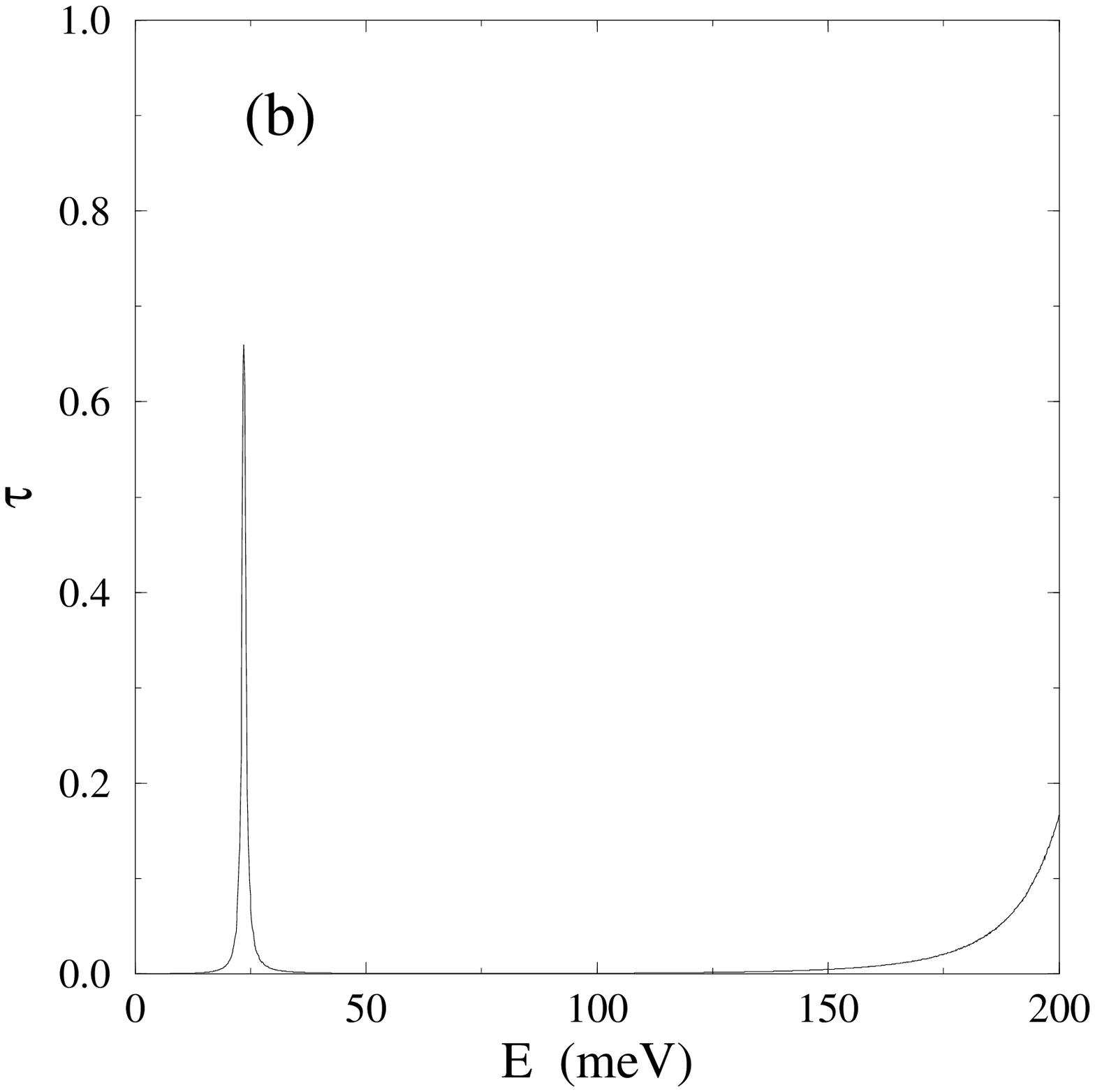}}}
\setlength{\epsfxsize}{3.8cm}
\centerline{\mbox{\epsffile{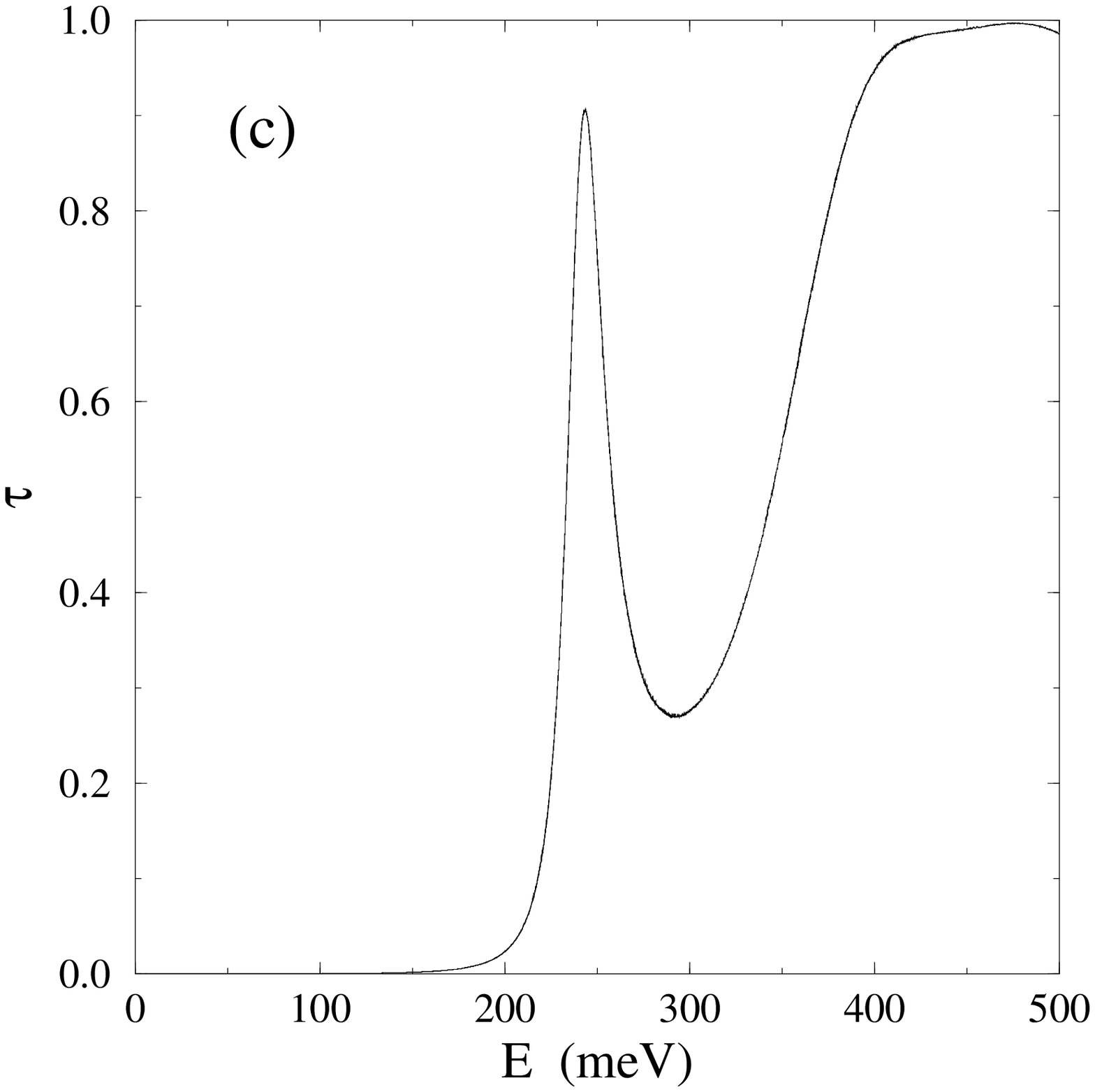}}}
\setlength{\epsfxsize}{3.8cm}
\centerline{\mbox{\epsffile{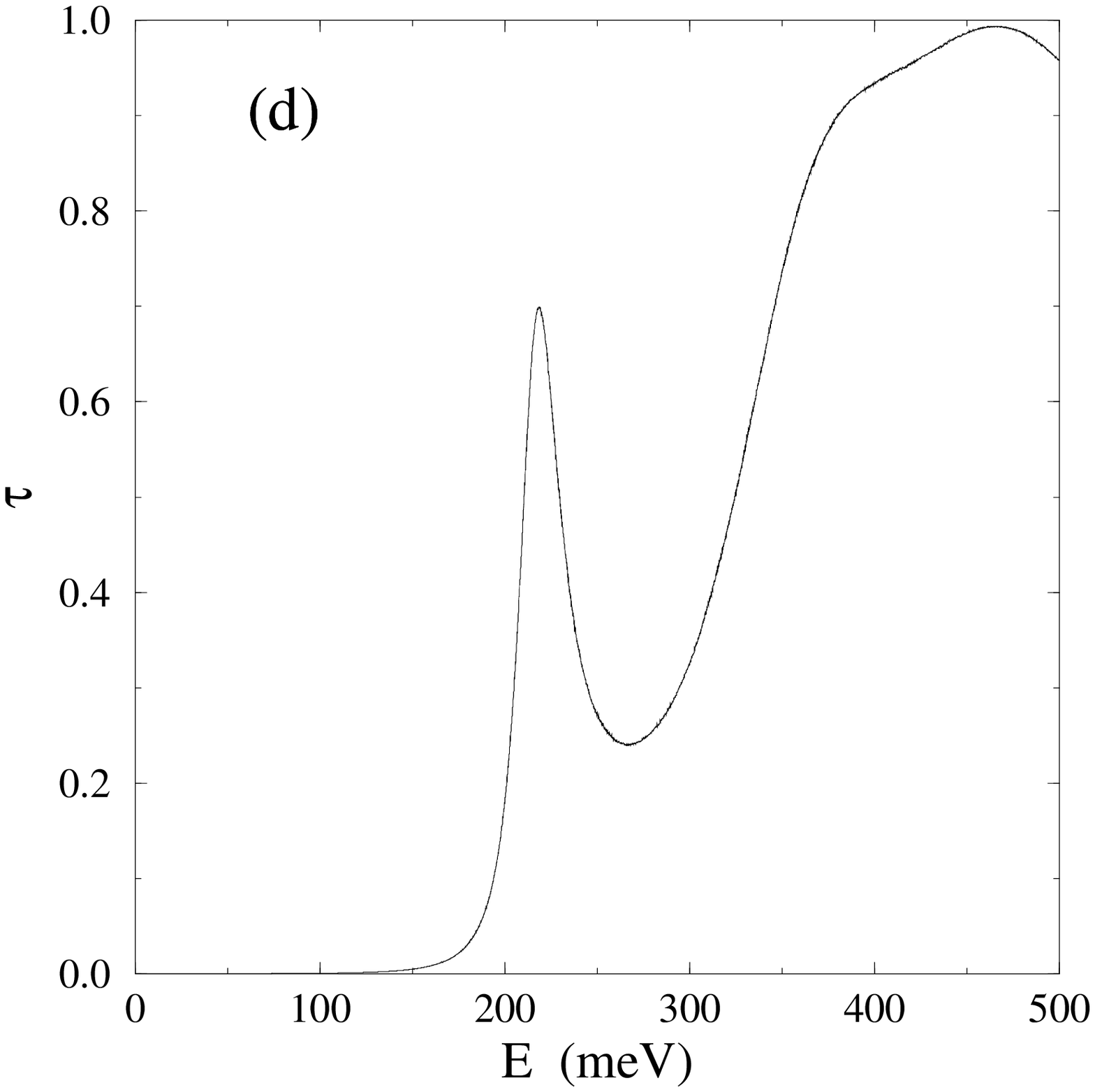}}}
\setlength{\epsfxsize}{3.8cm}
\centerline{\mbox{\epsffile{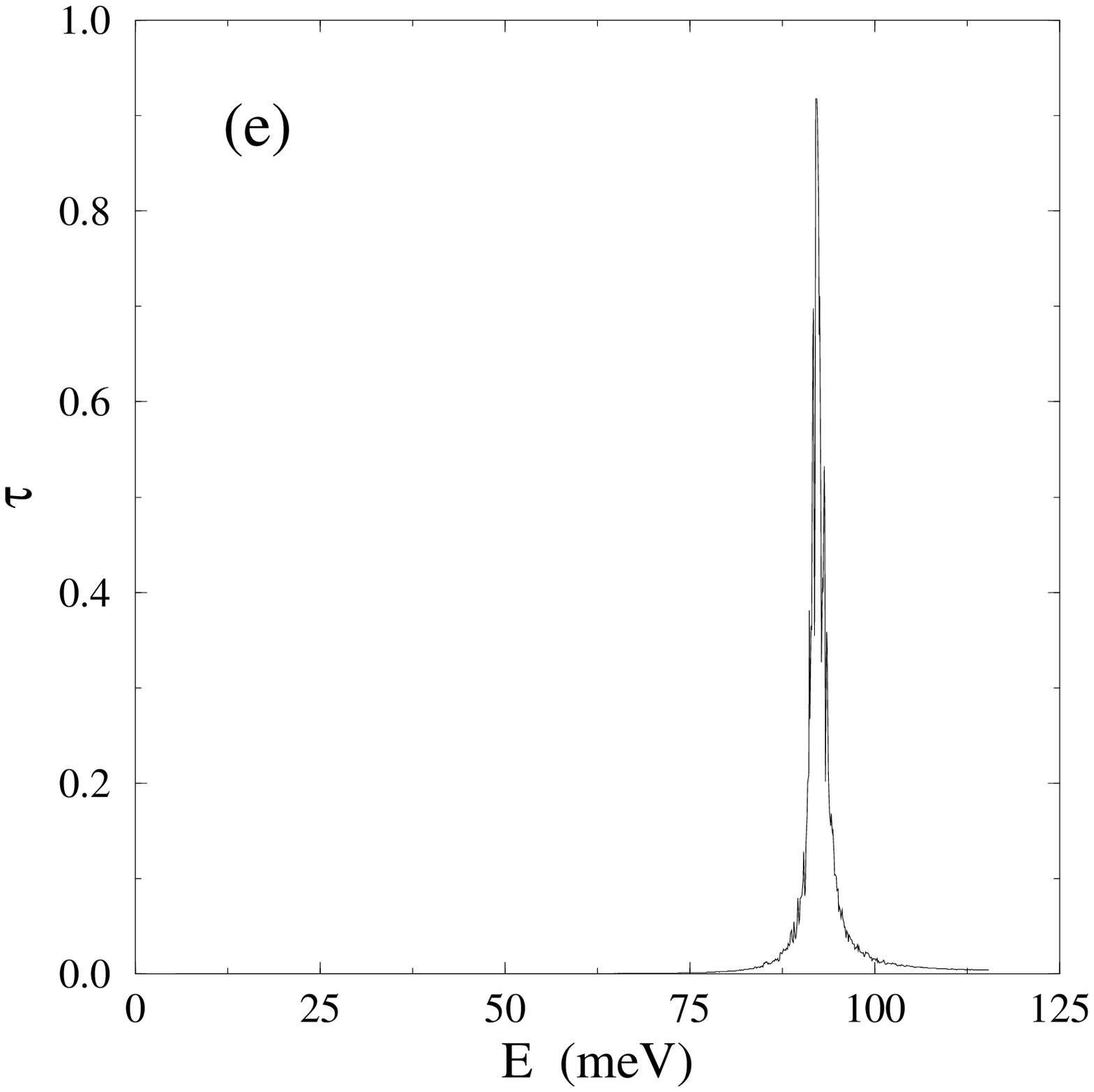}}}
\setlength{\epsfxsize}{3.8cm}
\centerline{\mbox{\epsffile{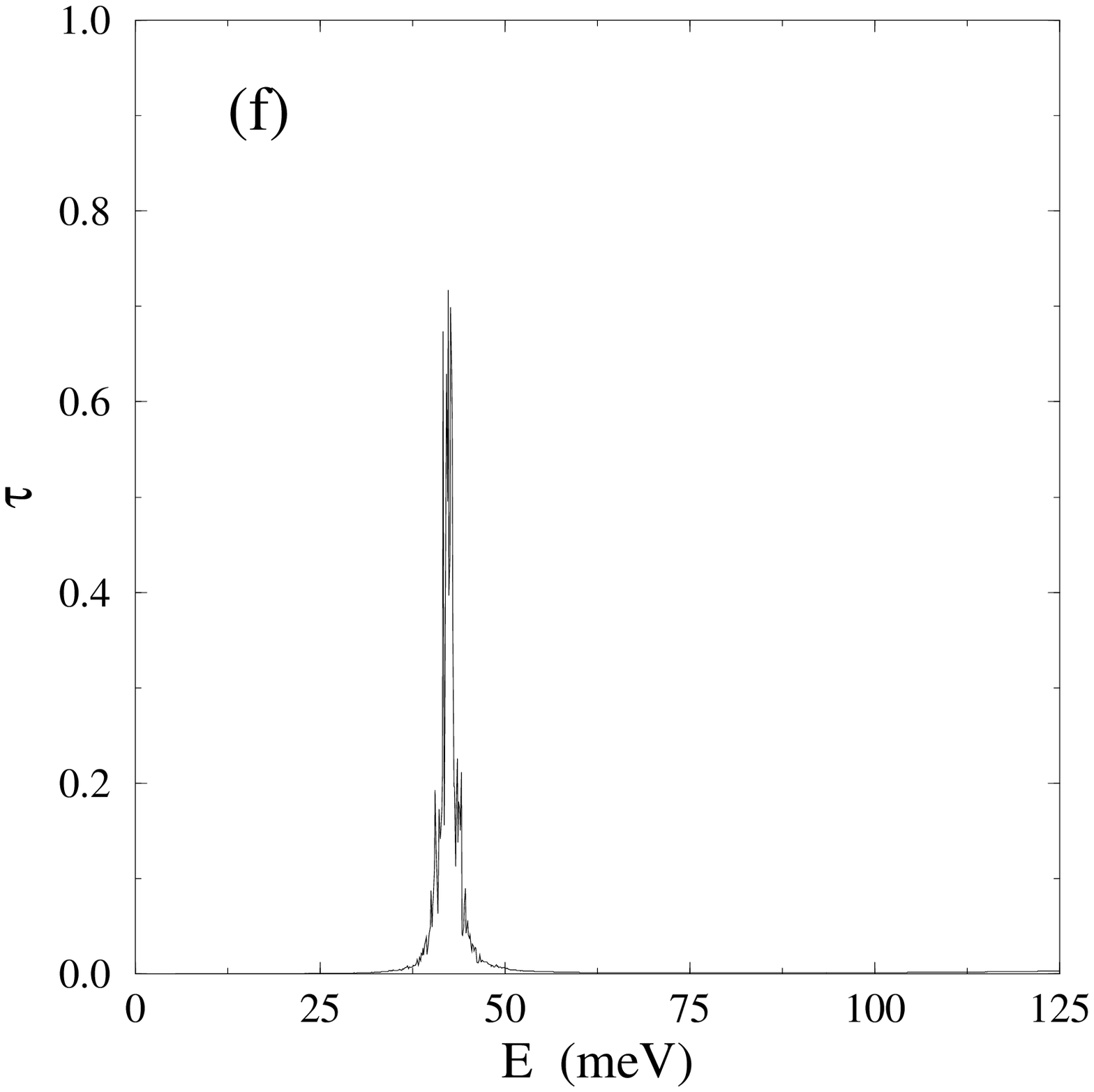}}}
\caption{Transmission coefficient $\tau$ as a function of the
electron energy for $\beta=-0.001$ with (a) $V=0.05\,$volts and (b)
$0.10\,$volts; $\beta=-0.01$ with (c) $V=0.05\,$volts and (d)
$0.10\,$volts; $\beta=-0.1$ with (e) $V=0.05\,$volts and (f)
$0.10\,$volts.}
\label{fig6}
\end{figure}

Let us now turn our attention to the case when $\alpha=0$, i.e., only
the well in the DBS is nonlinear, which turns to be out very different
from the previously discussed nonlinear barrier DBS. Results for the
transmission coefficient at zero field are shown in Fig.~\ref{fig4}.
{}From those plots, it can be seen that in general the effect of $\beta$
is more dramatic than that of $\alpha$, and that the case when
$\beta=-0.001$ has not much to do with the other two values.  For the
smallest nonlinearity, we see two peaks (as in the intermediate $\alpha$
case), but the main one has departed largely from the linear resonance.
This is related to the fact that $\beta$ is negative, and therefore,
accumulation of charge inside the well makes it deeper, thus moving the
resonant quasilevel to lower energies.  Increasing $\beta$ will
eventually locate the resonance below the zero energy and thus the
appearance of the plots for $\beta=-0.01$ and $\beta=-0.1$.  The less
pronounced peaks appearing (meaning, for $\beta=-0.001$, the one playing
the role of the sideband) have to do with the deformation of the bottom
of the well by the spatial distribution of the charge.  This can again
be understood from the same effective potential ideas discussed in the
previous subsection: By inspecting Fig.~\ref{fig5}, where the effective
potential is shown for the intermediate and large $\beta$ values, it is
clear that tunneling of electrons inside the well changes its structure
dramatically, at first even splitting it into two [Fig.\ \ref{fig5}(a)]
or several [Fig.\ \ref{fig5}(b)] wells.
The main point in that picture
is that for the largest nonlinearity, the depth of the well is much
larger than in the linear case, and therefore the main resonance is now
located at negative energies, this being the reason for the
disappearance of the main peak in the transmission coefficient.
Besides, the complicated structure of the bottom of the well in [Fig.\
\ref{fig5}(b)] helps understand why the peak appearing in the
transmission coefficient exhibits structure; coupling between the
subwells induces splitting of the quasi-level into several subresonances
responsible for this finest features of the transmission coefficient [see the
inset of
Fig.~\ref{fig4}(c)].
When bias is applied to the slightly nonlinear well ($\beta=-0.001$), the
structure of the transmission coefficient changes very smoothly, see
{}Fig.~\ref{fig6}, much as in the nonlinear barrier case.  Due to the
increased presence of electrons in the well induced by the field, the
resonant level goes down and so does the transmission coefficient peak.
Remarkably, the intermediate nonlinearity ($\beta=-0.01$) is such that there is
no peak
at all in the transmission coefficient, although for the largest field
value it is seen that one is entering from above.  This
will have consequences on the $j-V$ curves which will be discussed
below.

\begin{figure}
\setlength{\epsfxsize}{4.5cm}
\centerline{\mbox{\epsffile{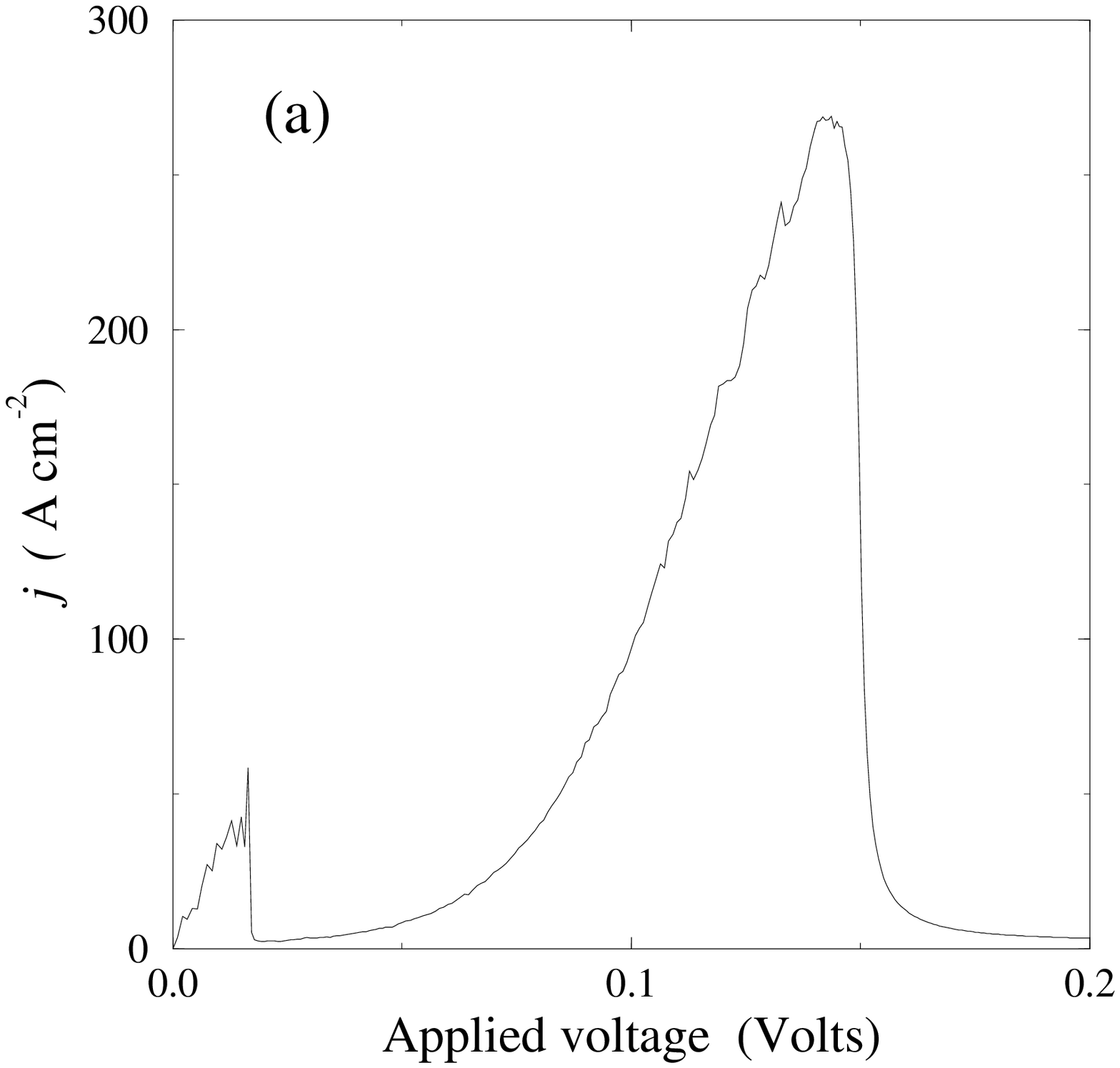}}}
\setlength{\epsfxsize}{4.5cm}
\centerline{\mbox{\epsffile{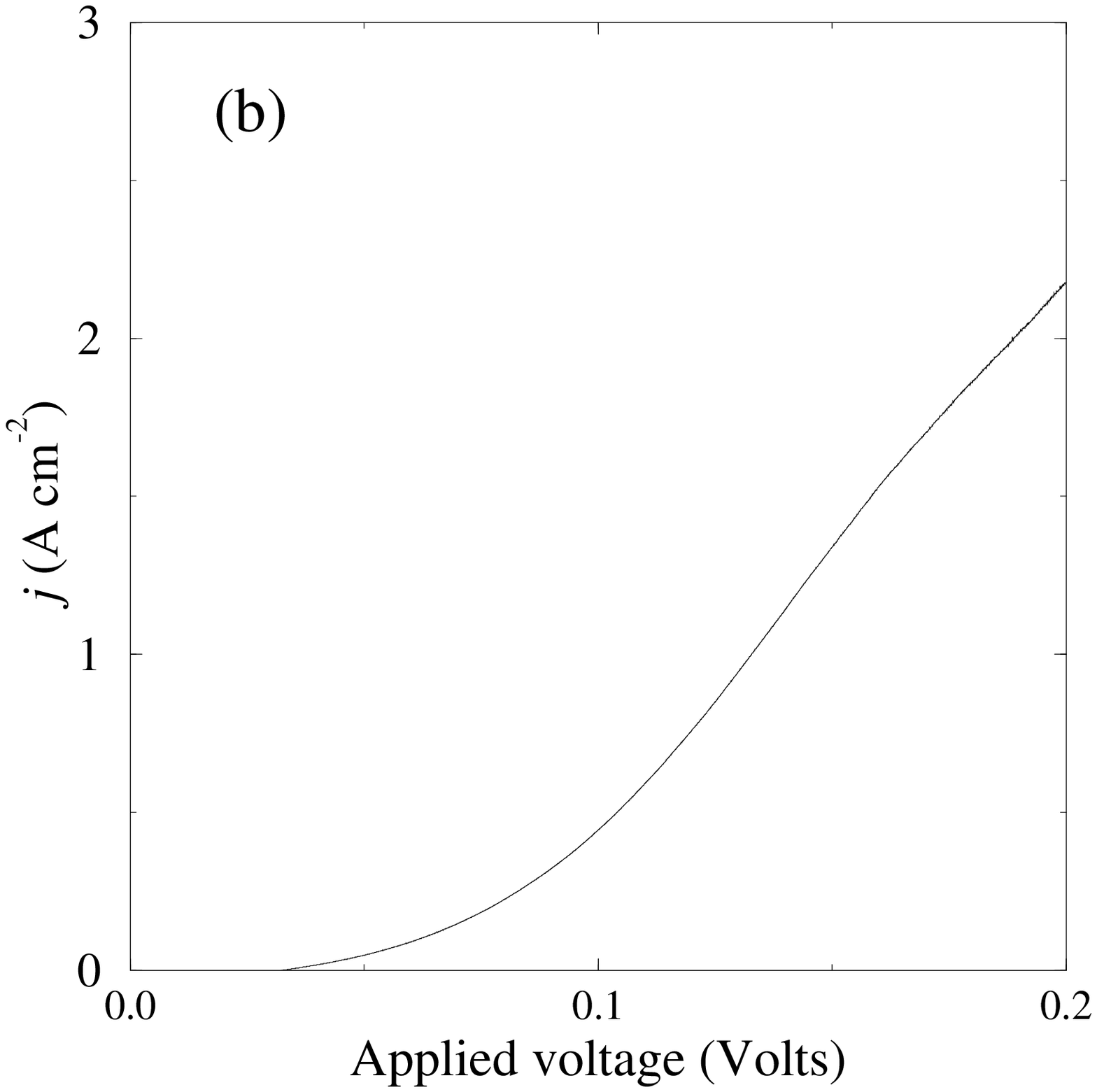}}}
\setlength{\epsfxsize}{4.5cm}
\centerline{\mbox{\epsffile{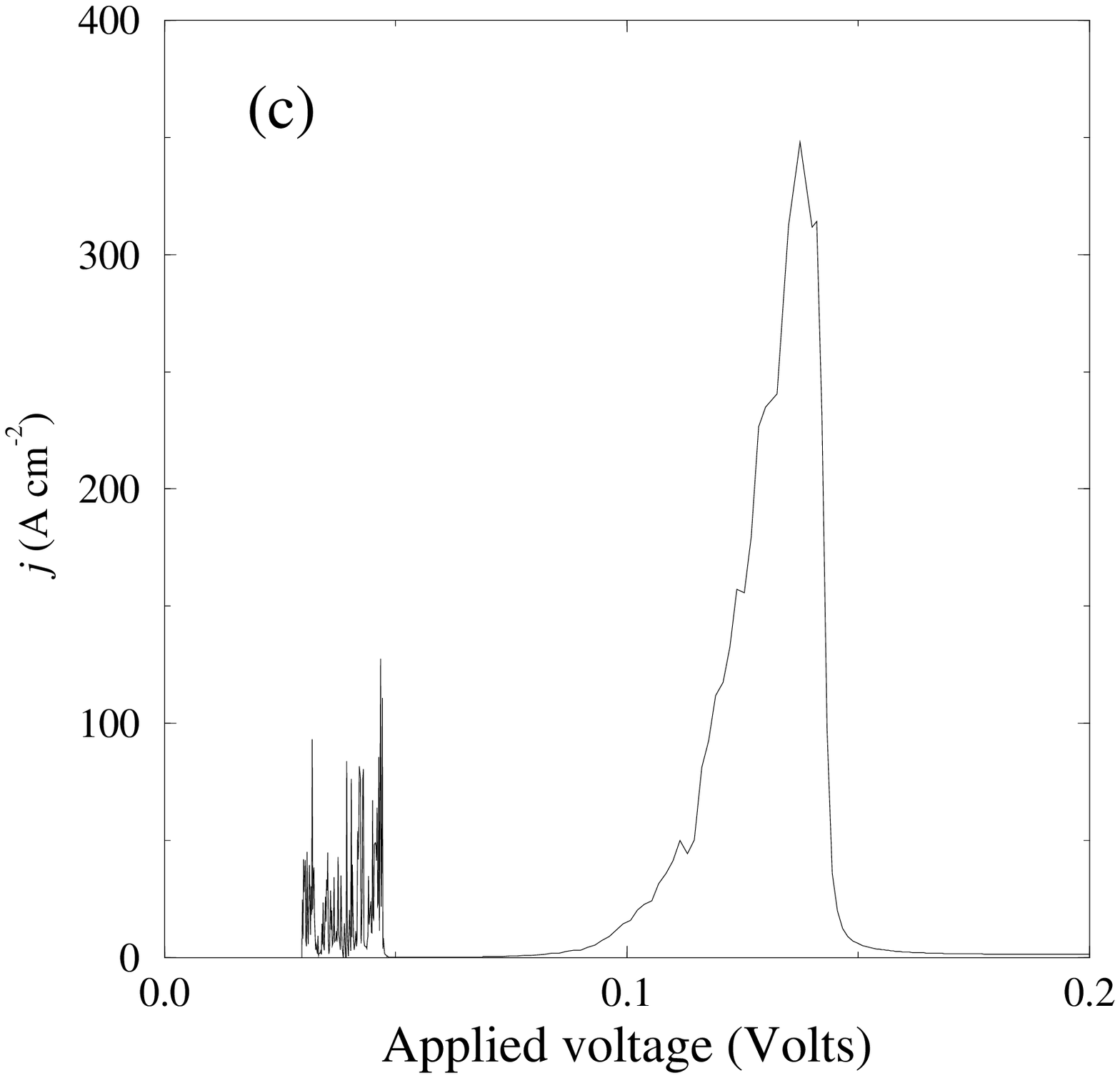}}}
\caption{Computed $j-V$ characteristics for $E_F=27.5\,$meV,
$T=77\,$K and (a) $\beta=-0.001$, (b) $-0.01$ , and (c) $-0.1$. Note that the
scale
in (b) is much smaller than in the other two ones.
Results at room temperature at the same although scaled by one half.}
\label{fig7}
\end{figure}

For $\beta=-0.1$, the field increases the inner structure of the
peaks but for the rest they behave as in the other cases, moving
downwards when increasing the field.  In view of the effective potential
structure, this probably has to do with the fact that when the field is
applied the tilting of the well induces the appearance of a positive
energy quasi-level, whose energy decreases with increasing field and,
subsequently, with increasing well depth.  Larger fields destroy this
structure, suppressing this quasi-level.  The different behavior for the
values of $\beta$ considered reflects in the $j-V$ characteristics as
well.  We only show the results at T$=77\,$K because those at room
temperature are essentially the same, scaled down by a factor one half.
In Fig.\ \ref{fig7}(a) it can be seen that the case $\beta=-0.001$ is
practically identical to the $\alpha=-0.01$ case, which is reasonable in
view that they both have the same transmission coefficient structure,
albeit for different reasons.  The main differences between those two
devices are that for this nonlinear well one we are discussing, the
first peak appears for much smaller values of the voltage, and that it
shows a lot of structure coming from the inner wells induced in the quantum
well.
Upon increasing the value of $\beta$, it is found that,
interestingly, intermediate $\beta$ values suppress completely the
current [Fig.\ \ref{fig7}(b)], a phenomenon that is evidently connected with
the absence of
resonances for that $\beta$ value already reported.  However, the case
of large $\beta$ shows a very unexpected feature: A series of noisy
peaks appears when the nonlinear DBS starts conducing, which we
tentatively associate to the structure of the well in this highly
nonlinear case.  For the rest, the DBS behaves qualitatively as in the
linear case, as seen in Fig.~\ref{fig7}(c).

\subsection{Fully nonlinear DBS}

When both constituents of the DBS, i.e., the barriers and the well, are
nonlinear, the number of possibilities is of course considerably
augmented.  It is not our aim to study thoroughly all possible
combinations in this section, but rather, to present a brief idea of
what can be said about this more complex devices.
\begin{figure}
\setlength{\epsfxsize}{3.9cm}
\centerline{\mbox{\epsffile{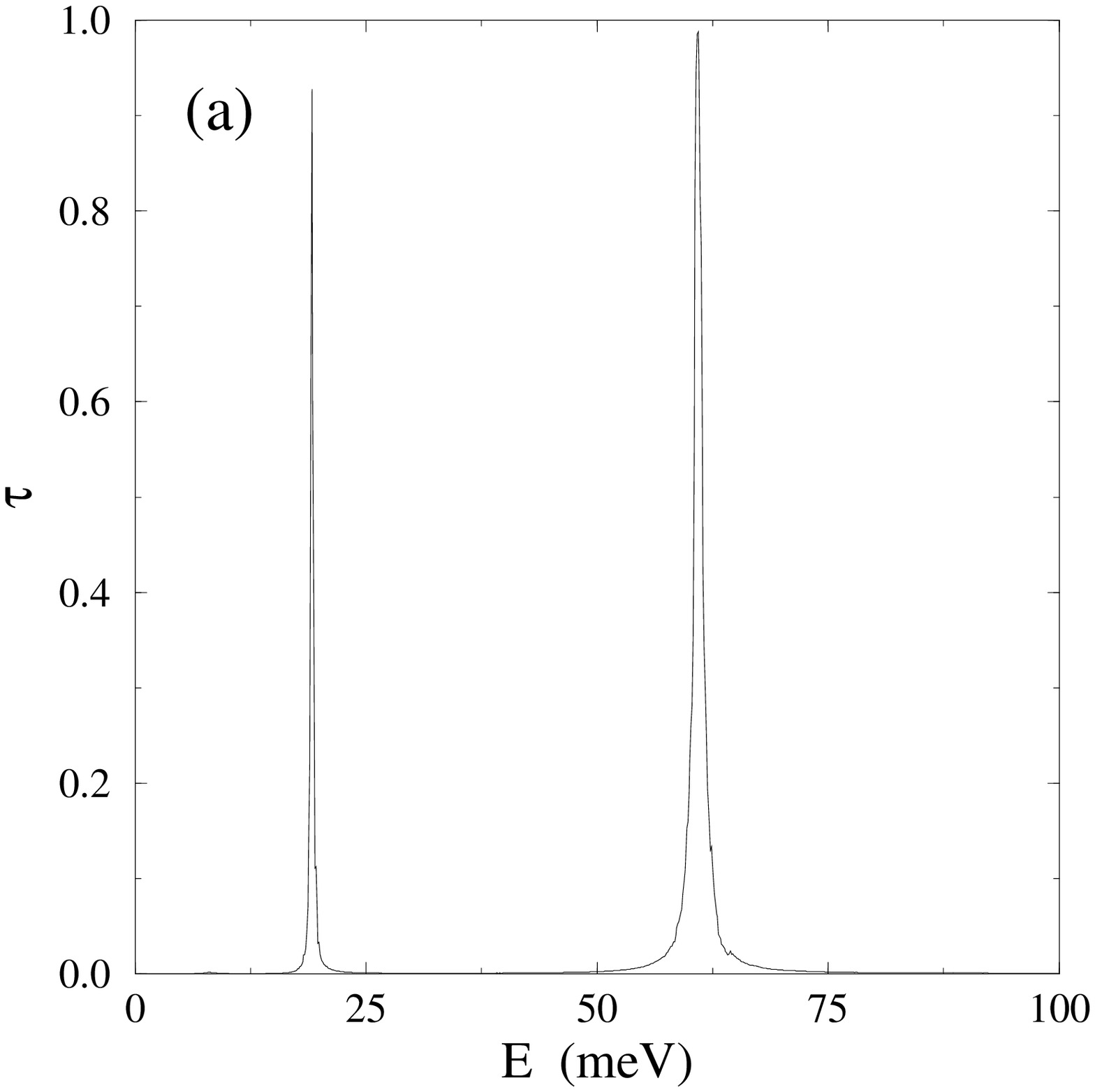}}}
\setlength{\epsfxsize}{3.9cm}
\centerline{\mbox{\epsffile{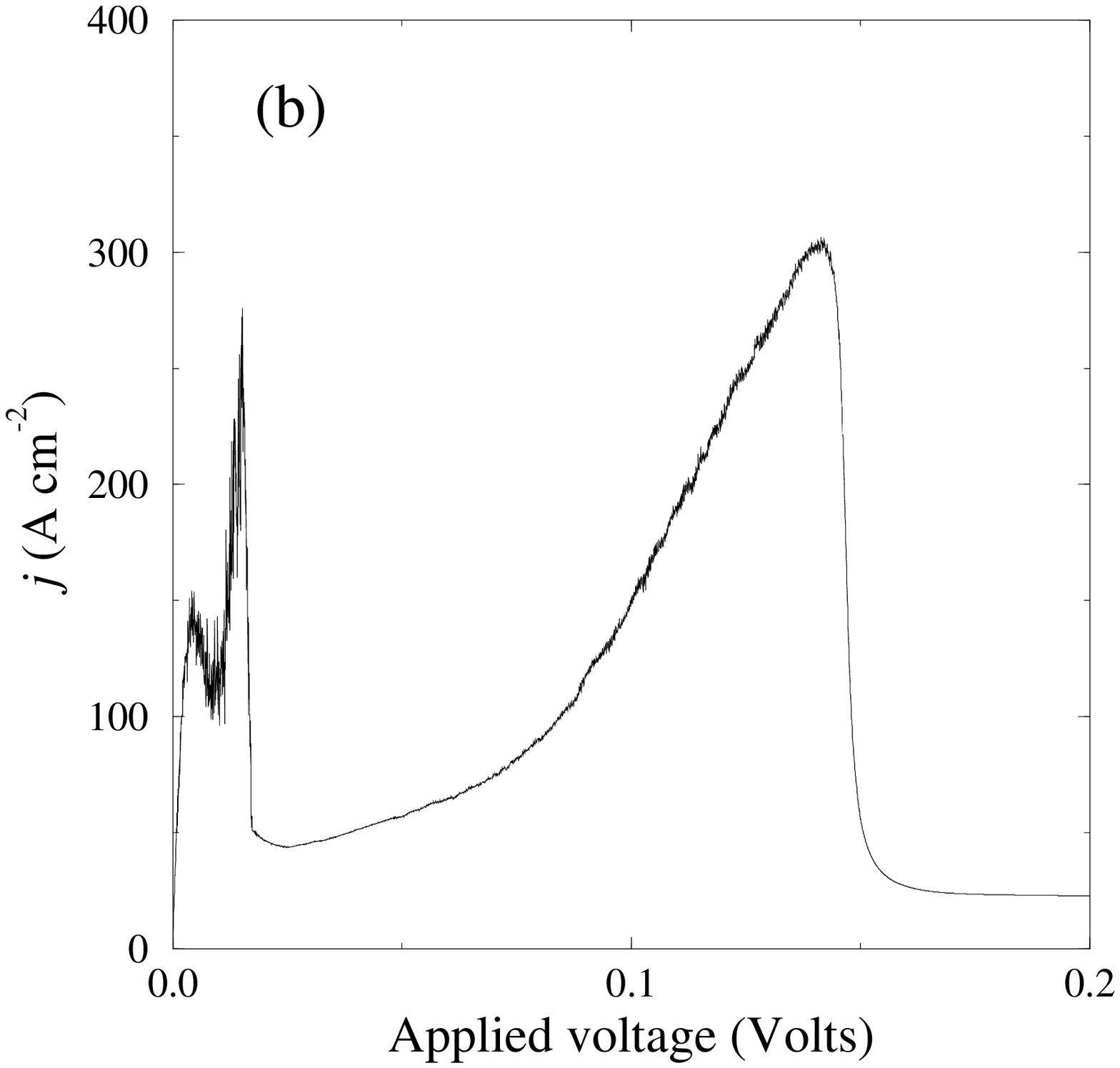}}}
\setlength{\epsfxsize}{3.9cm}
\centerline{\mbox{\epsffile{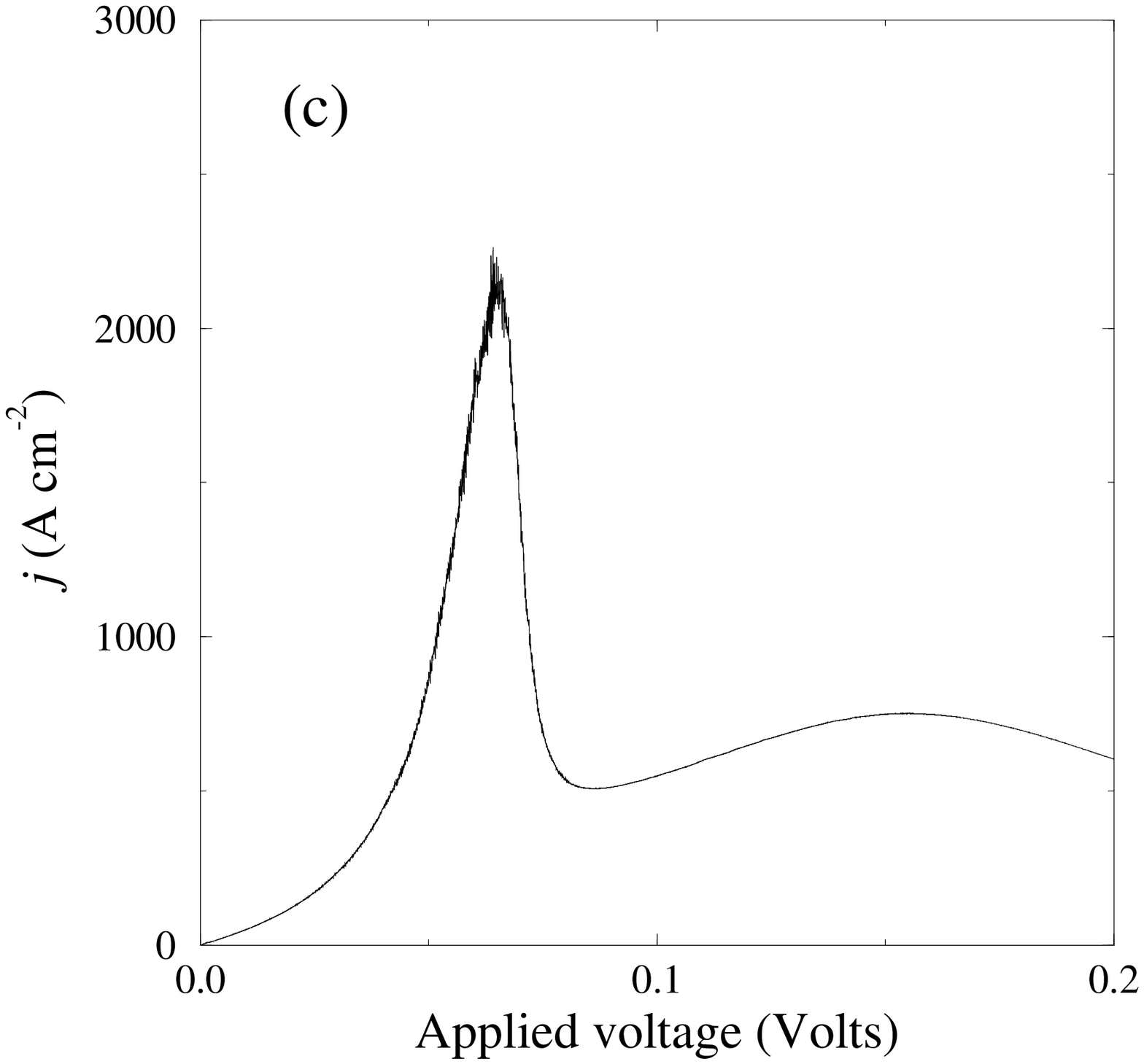}}}
\caption{(a) Transmission coefficient $\tau$ as a function of the
electron energy for $\alpha=\beta=-0.001$. (b) $j-V$ curve for
$\alpha=10\beta=-0.01$ and $T=77$ K. (c) $j-V$ curve for
$\alpha=\beta=-0.1$ and $T=77$ K.}
\label{fig8}
\end{figure}
After an exploration
of the main possibilities, we concluded that for small and intermediate
$\alpha$ (barrier nonlinearity), it is the nonlinearity of the well,
$\beta$, the one that governs the behavior of the system.
Thus, for
$\beta=-0.001$, the structure of the transmission coefficient shows two
peaks until $\alpha=-0.1$ which leads to a smoothing out of the peaks by
almost suppression of the barriers.  Indeed, the two peak structure is
even clearer when $\beta=\alpha=-0.001$ than in the other cases
discussed so far [see Fig.~\ref{fig8}(a)], and as consequence so is the
two NDR peak feature of the $j-V$ curve.  The only difference between
$\alpha=-0.001$ and $\alpha=-0.01$ is that this last choice always gives
rise to a somewhat more complicated curve [see an example in Fig.\
\ref{fig8}(b)].  On the other hand, when $\alpha$ is large, then it is
this factor the one responsible for the behavior of the DBS, except when
$\beta$ is also large (thus compensating the decreasing of the
barriers).  In that case, as shown in Fig.\ \ref{fig8}(c), it is
possible to obtain a device capable of supporting a high current while
still keeping a neatly marked NDR peak.  We thus see that the desired
characteristics of devices consisting only of one nonlinear component
can be fine tuned by introducing nonlinearity in the other component, or
improved by suitably combining both.

\section{Conclusions}

In this concluding section of the paper, we summarize briefly what are
our main findings on the transport properties of nonlinear DBS. A first
overall conclusion is that the behavior of such kind of devices changes
quite a bit when the nonlinearity of the system changes.  This is very
important: Keeping in mind that the nonlinear coefficients we have been
handling include the contribution of the amplitude of the wavefunction,
it is immediately understood that a device, built up from specific
materials with specific and constant nonlinear characteristics, will
change dramatically its response if the amplitude of the incoming
wavefunction is also modified.  This in turn can be related to the
incoming current density.  Clearly, a result like this paves the way to
the search for devices with particular features arising from
nonlinearity.  We have to remind here that even if we have been
discussing the outcome of our work in terms of electron transport, it
can be straightforwardly translated to optical contexts, which gives
further relevance to our results in view of the increasing importance of
nonlinear optical phenomena for communications.  In this respect, we
want to stress that for that application the control of the nonlinearity
of the devices by the incoming wave is even easier to realize.

Being more specific, we have found that devices with nonlinear barriers
and negative, i.e., self-attractive nonlinearity, definitely show an
increase of the current as compared to the linear case.  This fact is
not strange once it is realized that the role of nonlinearity in this
case is to reduce the barriers.  Roughly speaking, nonlinear barrier DBS
begin transporting current for lower voltages; afterwards, they exhibit
constant or slowly increasing characteristics for a range of voltages,
which ends up in a NDR region.  The features of the $j-V$ curves are of
course smoothed when the temperature is rised up to room values, a
conclusion which applies to all the nonlinearities considered in this
work.  On the other hand, if the DBS is equipped with a nonlinear well,
the device has more striking properties: In the lowest nonlinearity
situation, it exhibits highly nonlinear properties, such as two NDR
peaks; intermediate values of the nonlinear contribution will strongly
suppress the current, and, finally, high nonlinearities lead to a
quasi-linear current-voltage characteristic, with the RT peak being
shifted to lower voltages and the current slightly increased.  This is a
peculiar property, and therefore we want to draw the readers' attention
to the curious fact that nonlinear well DBS behave more differently from
the linear case in the small nonlinearity situation.  If we take also
into account the suppression of the current for intermediate nonlinear
values, it is evident that this kind of devices is the most appealing
candidate for new applications.  Finally, we have briefly pointed out
some examples of combining both nonlinearities, and we have seen that it
is possible to obtain clearer NDR peaks or higher conductivities (or
both) by appropriately choosing the two components of the DBS to be
nonlinear.

In view of this report, we feel we can conjecture than complex
structures consisting of a number of nonlinear DBS will also exhibit
very peculiar properties.  One can think, for instance, of a system
built up of nonlinear well DBS. If the input to such device is of high
amplitude, the first DBS components will behave quasi-linearly, letting
the current pass through them.  The deviation from unity of the
transmission coefficient will make decrease the amplitude, thus reaching
the region for which the current is severely suppressed.  This, in turn,
would give rise to current in the remainder of the structure, because
then it would behave more nonlinearly (as discussed above) allowing for
transmission of the wavepacket.  We would have thus designed a device
which would transport current differently as a function of its length.
Another possibly relevant remark relates to the mutual influence of
disorder and nonlinearity, and applies to the case of nonlinear barrier
DBS. It is clear that if a superlattice made of linear DBS is built, the
imperfections introduced during the growth will lead to a mismatch among
the resonant levels of neighboring wells and as consequence to a loss of
quantum coherence leading to poor transport properties.  Nonlinear
barrier DBS will be much more robust against growth imperfections, as
they have a plateau-like current behavior for a range of voltages, which
makes irrelevant the exact matching of quasi-levels.  The conclusion is
once more that nonlinearity would help transport even in the presence of
disorder.  In any event, a detailed study of more complex structures is
necessary to clarify and put on firm grounds all these ideas.

We finish by briefly pointing out another group of open questions which
arise in view of our work.  Further extensions of the present work to
study nonlinear dynamical response of DBS on external ac bias would be
of great interest to shed light on related problems like bistability,
\cite{Pawel} noise characteristics, \cite{Flores} and RT at far
infra-red frequencies \cite{Chitta} under the influence of inelastic
scattering channels as those described here.  Besides that, the above
mentioned highly nonlinear limit could also be interesting, for
nonlinearity is always susceptible to give rise to new and unexpected
features.  If these new features were seen in our model it would be a
very exciting development.  If materials with suitable characteristics
were found, and that is to be expected, experiments could be made to
check the predictions: If the model were wrong, that would establish its
range of validity, whereas if the predictions were correct, this work
could pave the way to a new family of devices and applications.

\acknowledgments

We are very thankful to Paco Padilla for useful discussions.  A.\ S.\
acknowledges partial support from C.I.C.\ y T.\ (Spain) through project
No.\ PB92-0248 and by the European Union Human Capital and Mobility
Programme through contract ERBCHRXCT930413.

\appendix

\section*{Nonlinearity sign}

In this appendix, we give mathematical reasons why $\alpha$ and $\beta$
(or $\tilde{\alpha}$ or $\tilde{\beta}$) should be negative by using
Eq.~(\ref{q}).  We focus in the study of the problem with
$\tilde{\beta}=0$ for simplicity, but similar considerations apply to
the general case.  With this assumption, the equation has the form
\begin{equation}
\label{q2}
-q_{zz}(z)+{1\over q^3(z)}+f_1(z)q(z)+f_2(z)q^3(z)=0,
\end{equation}
where $f_1(z)$ and $f_2(z)$ are well-behaved functions.  Let us now
consider the discrete version of this equation, with the second
derivative discretized in the usual way; denoting $q_n=q(z=n\Delta z)$
and $f_{in}=f_i(n\Delta z)$, $i=1,2$, with $\Delta z$ being the
integration step, Eq.~(\ref{q2}) can be rewritten as
\begin{equation}
\label{q3}
q_{n+1} = 2q_n - q_{n-1}+(\Delta z)^2 \left[ {1\over q_n^3} + f_{1n} q_n
+ f_{2n}q_n^3\right].
\end{equation}

Notice that in this expression the sign of $f_{2n}$ is the same as the
sign of $\alpha$ and $\tilde{\alpha}$.  Let us now consider
Eq.~(\ref{q3}) for large $q_n$; this is general, because if $q$ is small
the term $q_n^{-3}$ will make it grow quite quickly.  In this limit,
Eq.~(\ref{q3}) can be approximately replaced by
\begin{equation}
\label{q5}
\Delta q_{n-1} = \Delta q_n + (\Delta z)^2 f_{2n}q_n^3,
\end{equation}
where $\Delta q_n=q_n-q_{n+1}$, and we have cast the equation in this fashion
because it is to be integrated backwards. Recalling the initial conditions
(\ref{ic}), if $N$ is the total number of grid points, we have $q_N=
q_{N-1}=k_L^{-1/2}>0$. Therefore,
we see that $\Delta q_{N-1}=0$, and
\begin{equation}
\label{eh}
\Delta q_{N-2}=(\Delta z)^2 f_{2n}
k_L^{-3/2}.
\end{equation}
We thus see that if $f_{2n}$ is positive (and hence $\alpha$) the first
increment is positive, and so are the subsequent ones, leading to a
exponential divergence of $q$, whereas if $f_{2n}$ and $\alpha$ are
negative, the increment is negative, and $q$ decreases until the
$q_n^{-3}$ starts being relevant again.  In this last case, it is
possible that $q$ reaches an equilibrium due to the balance of the two
cubic terms, which was not for $\alpha>0$.  This is in fact seen in the
numerical integration of Eq.~(\ref{q}), where $q$ rapidly diverges if
$\alpha$ is positive, unless, of course, $\alpha$ is positive but very
small and/or the barrier (the region for $\alpha$ to influence $q$) is
very narrow.  In this last situation, however, the effect of $\alpha$
becomes negligible.

Physically, this can be understood as follows.  Seeing $z$ as a time
variable, Eq.\ (\ref{q2}) can be regarded, loosely speaking, as an
evolution equation for $q$, which is equivalent to the following
interpretation: Electrons impinge on the barrier from outside, and begin
to tunnel through it, their wavefunction being real and exponentially
increasing or decreasing.  If $\alpha$ is positive, then if there were
any charge density in the BDS, this would become even more repulsive,
and the wavefunction would diverge even faster (numerically one always
see the exponentially growing part, of course).  This instability is not
present in the opposite case, where a negative $\alpha$ helps the
electron tunnel across the barrier.  Note that this reasoning does not
apply to the full partial differential equation, which is the one we
should deal with for that case by starting from the complete
Schr\"odinger equation.

\end{multicols}

\end{document}